\newcommand\astroph[2]{preprint (astro-ph/#2)}

\documentclass[12pt,preprint,eqsecnum]{aastex}
\usepackage{epsfig}

\begin{document}

\title{ASYMMETRIC BEAMS IN COSMIC MICROWAVE BACKGROUND
 ANISOTROPY EXPERIMENTS}

\author {J.~H.~P.~Wu\altaffilmark{1},
  A.~Balbi\altaffilmark{2,3,4}, 
  J.~Borrill\altaffilmark{5,3},
  P.~G.~Ferreira\altaffilmark{6,7},
  S.~Hanany\altaffilmark{8,3},
  A.~H.~Jaffe\altaffilmark{3,9,1},
  A.~T.~Lee\altaffilmark{10,3,9},
  S.~Oh\altaffilmark{3,10},
  B.~Rabii\altaffilmark{3,10},
  P.~L.~Richards\altaffilmark{10,3},
  G.~F.~Smoot\altaffilmark{10,3,4,9},
  R.~Stompor\altaffilmark{3,9,11},
  C.~D.~Winant\altaffilmark{3,10}
}

\altaffiltext{1}{Dept. of Astronomy, University of California,
  Berkeley, CA, USA}

\altaffiltext{2}{Dipartimento di Fisica, Universit\`a Tor Vergata,
  Roma, Italy}

\altaffiltext{3}{Center for Particle Astrophysics, University of
  California, Berkeley, CA, USA}

\altaffiltext{4}{Lawrence Berkeley National Laboratory, University of
  California, Berkeley, CA, USA}

\altaffiltext{5}{National Energy Research Scientific Computing Center,
  Lawrence Berkeley National Laboratory, Berkeley, CA, USA}

\altaffiltext{6}{Astrophysics, University of Oxford, UK}

\altaffiltext{7}{CENTRA, Instituto Superior Tecnico, Lisboa,
  Portugal}

\altaffiltext{8}{School of Physics and Astronomy, University of
  Minnesota/Twin Cities, Minneapolis, MN, USA}

\altaffiltext{9}{Space Sciences Laboratory, University of California,
  Berkeley, CA, USA}

\altaffiltext{10}{Dept. of Physics, University of California,
  Berkeley, CA, USA}

\altaffiltext{11}{Copernicus Astronomical Center, Warszawa, Poland}


\begin{abstract}
We propose a new formalism to handle asymmetric beams in the data
analysis of cosmic microwave background anisotropy experiments. 
For any beam shape,
the formalism finds the optimal circularly symmetric equivalent
and is thus easily adaptable to existing data analysis methods.  
We demonstrate certain key points
by using a simulated highly elliptic beam,
and the beams and data of the MAXIMA-1 experiment,
where the asymmetry is mild.
We show that in both cases 
the formalism does not bias the angular power spectrum estimates.
We analyze the limitations of the formalism
and find that
it is well suited for most practical situations.

\end{abstract}


\section{INTRODUCTION}
\label{introduction}

A new generation of Cosmic Microwave Background (CMB) mapping experiments is
beginning to produce data of unprecedented quality (see 
e.g., Torbet et al.~1999; Miller et al.~1999; 
DeBernardis et al.~2000; Hanany et al.~2000, hereafter H00).
Much of the experimental effort is concentrated on probing angular
scales of about 10 arcminutes. 
To fully benefit
from the scientific potential of  these  high resolution data sets,
a new level of sophistication is required in quantifying all 
possible sources of error in the experimental procedure and
data analysis pipeline 
(e.g., Ferreira \& Jaffe 2000). 
Particular care must be used to accurately quantify the 
instrument response to the signal  
and to include such response in the data analysis.

In all analyses of CMB data so far the experimental beam has
been assumed to have a radial symmetry. This assumption has been
incorporated in most map-making and angular power spectrum ($C_\ell$) estimation
algorithms (e.g., Bond, Jaffe, \& Knox 1998) and is necessary because of
limitations in computing capability.  A crude
symmetric-beam approximation was adequate in the past since most of
the error budget was dominated by statistical and other systematic
uncertainties.  However, with the precision of current and future
analyses, it becomes essential to establish a methodology for
accurately quantifying the degree of beam-asymmetry and properly 
incorporating it
into the data analysis pipeline.  If the beam is incorrectly
incorporated in the data analysis pipeline, one may not only
artificially distort the underlying structure of the measured CMB
signal but also bias the estimate of the CMB angular power spectrum.
In this paper we present a new formalism for estimating the 
power spectrum that can handle any beam shape. 
We show that the formalism can be applied to a broad variety of cases
which encompass most practical applications. As a consequence, the detailed
shape of the antenna beam should no longer pose a limitation in 
measuring the angular power spectrum of CMB experiments.

The asymmetry of beams may arise from a variety of sources.  For
example, it may be due to the optics, or due to the finite response
time of a detector which leaves imprints in the direction of the scan
(e.g., Hanany, Jaffe, \& Scannapieco 1998).  Regardless of the origins of
the asymmetry, the framework we shall present is general, and consists
of finding an equivalent symmetric beam that replaces the 
asymmetric beam in the analysis of the data.  

Using the formalism 
one can assess the degree of asymmetry of a beam (see eq.~[\ref{varpi_l}]),
how the asymmetry propagates through the analysis pipeline,
and
how to find an azimuthally symmetrized beam that best approximates the
asymmetric beam (see e.g., eq.~[\ref{Bl2eff_avg4}]).
The symmetrized beam is then used in the
symmetric-beam approximation of the $C_\ell$ estimation 
(see eq.~[\ref{CBBB_app3}]). 
The formalism quantifies 
the errors introduced in the $C_\ell$ estimates because of
the use of the symmetrized-beam approximation, 
the uncertainty in the final $C_\ell$ estimates
resulting from the uncertainty in the beam measurement 
(see eq.~[\ref{Delta_Cl}]), 
and 
the smoothing effects due to the pixelization
of the map (see eqs.~[\ref{Bl2_effx}] and [\ref{Bl2_effxapp}]).  
It also shows 
how to combine beams from independent experimental photometers (see
eqs.~[\ref{W_l}], [\ref{barBplm2}], [\ref{chi_t}], and [\ref{mu_t}]).
Some useful conditions under
which this new formalism will be needed are also provided 
(see eqs.~[\ref{Clmid_constraint}] and [\ref{Cl2mid_constraint}]).

The structure of this paper is as follows.
In section~\ref{the_convention},
we describe the framework of CMB data analysis
for the estimation of the power spectrum,
so as to illustrate the problems related to asymmetric beams.
In section~\ref{the_criteria_for_beam_symmetry},
we define 
the `index of asymmetry' (IOA) $\varpi_\ell$,
a useful parameter in quantifying the level of asymmetry of a beam.
Similarly,
we define the `index of combined asymmetry' (IOCA) ${\cal W}_\ell$,
which is useful when combining data 
from photometers of different beam shapes.
In sections~\ref{the_effective_Bl_of_pixels} and
\ref{the_effective_two_point_Bl},
we investigate the problems associated with asymmetric beams.
We introduce
the `average pixel-beam expansion', $\overline{B}_{p\ell m}$, and
the `pixel-pixel beam expansion', $B_{\ell{\rm (eff)}}^2$,
to
provide an approximation scheme
where the convolution effect of asymmetric beams
is treated as circularly symmetric.
The biasing effects of this approximation
in the resulting estimated power spectrum $C_\ell$ are also considered.
In section~\ref{symmetry_vs_asymmetry},
we derive the conditions 
under which one needs to employ the new formalism 
for treating asymmetric beams.
In section~\ref{uncertainties_from_beam_measurement},
we investigate the uncertainties in the $C_\ell$ estimates resulting from
the uncertainties in the measurement of beam shape.
In section~\ref{deconvolution_of_the_pixel_smoothing},
we discuss another convolution effect due to the pixelization of
the CMB map. 
Although this is not a beam-related issue,
we demonstrate a simple way to incorporate its treatment into our framework.
In section~\ref{numerical_verifications},
we numerically verify
certain key points developed in sections~\ref{the_criteria_for_beam_symmetry}
to \ref{deconvolution_of_the_pixel_smoothing},
as well as the accuracy of the proposed approximation in treating
asymmetric beams.
In particular,
we use the data from the MAXIMA-1 experiment as an example
to demonstrate the generic treatment of asymmetric beams in CMB experiments.
It is shown that
our formalism has no biasing effects
in the resulting $C_\ell$ estimates.
Finally
in section~\ref{conclusion},
we summarize the procedure in applying our formalism to experiments,
discuss its availability,
and draw a conclusion.

\section{THE CONVENTION AND PROBLEMS}
\label{the_convention}

We first consider the standard procedure for the power spectrum estimation.
This consists of two main steps.
First, one estimates the pixelized map $m_p$ from a given time-stream $d_t$,
i.e., to translate the observation from the temporal ($t$) to
 the spatial ($p$) domain.
Second,
one estimates the power spectrum $C_\ell$ from the map $m_p$.

In the temporal domain,
what we observe is
\begin{equation}
  \label{d_t}
  d_t=\gamma_t + n_t,
\end{equation}
where
$\gamma_t$ is the CMB signal and
$n_t$ is the instrumental noise.
Traditionally we model the CMB observation as
\begin{equation}
  \label{gamma_t}
  \gamma_t=A_{tp}s_p,
\end{equation}
where we use the Einstein summation convention here and below when appropriate
(usually over pixels and time samples, but not over spherical harmonic indices).
Here $A_{tp}$ is the pointing matrix giving the weight of pixel
$p$ in observation $t$, and $s_p$ is the CMB signal on the pixel
convolved by a pixel beam $B_{p}({\bf x})$:
\begin{equation}
  \label{s_p}
  s_p=\sum_{\ell=0}^\infty\sum_{m=-\ell}^{\ell}
                   B_{p\ell m} a_{\ell m} Y_{\ell m}({\bf x}_p),
\end{equation}
where $Y_{\ell m}$ are the spherical harmonics,
and $B_{p\ell m}$ and $a_{\ell m}$ are the multipole expansions
of $B_p({\bf x})$ and the CMB signal respectively.
Note that we use a
two-dimensional vector ${\bf x}$ to denote locations on the surface of
the sphere, which we shall often consider in the small-field limit (see later).

 We usually take the pointing operator $A_{tp}$ to be one when
observing pixel $p$ at time $t$ and zero otherwise.
That is, we model the signal $\gamma_t$ to be the same for any observation
within pixel $p$. In effect, we take the sky to be smoothed with a
top-hat of shape given by the pixel boundary. 
We shall see in section~\ref{deconvolution_of_the_pixel_smoothing} that, as
expected, this is equivalent to an extra convolution included in $B_p$.

With this modeling, one can thus estimate the pixelized map from the
temporal data.  This involves maximizing the likelihood of the signal
given the data:
\begin{eqnarray}
  \label{L_s}
  {\cal L}(s) \propto {\rm Prob}[d|s] = 
  (2\pi)^{-{\cal N}_t/2} \times \nonumber\\
  \exp\left\{-\frac{1}{2}
  \left( n^T N^{-1} n + Tr[\ln N] \right)
  \right\},
\end{eqnarray}
where
$d\equiv d_t$, $s\equiv s_p$, and $n\equiv n_t$,
all as defined in equations (\ref{d_t}) and (\ref{gamma_t}),
${\cal N}_t$ is the size of the time-stream,
and
$N \equiv N_{tt'} = \langle n_t n_{t'}^T\rangle$ is
the time-time noise correlation matrix.
Here we have assumed
that the noise is Gaussian and
that all CMB maps are a priori equally likely.
Maximizing over $s$ gives
\begin{eqnarray}
  \label{m}
  m_p \equiv m &=& (A^T N^{-1} A)^{-1}(A^T N^{-1} d)\nonumber\\
&=& s_p + n_p, 
\end{eqnarray}
where $A\equiv A_{tp}$ as defined in equation (\ref{gamma_t}) and $n_p$
is the noise in the pixel domain.

One then moves on to estimate the power spectrum of the map,
$C_\ell=\langle |a_{\ell m}|^2\rangle$.
This requires the maximization of the likelihood function
\begin{eqnarray}
  \label{L}
  {\cal L} = {\rm Prob}[m|C_\ell] = 
  (2\pi)^{-{\cal N}_{C_\ell}/2} \times \nonumber\\
  \exp\left\{-\frac{1}{2} \left(m^T M^{-1} m + Tr[\ln M]\right)\right\},
\end{eqnarray}
where
${\cal N}_{C_\ell}$ is the dimension of the parameter space of $C_\ell$, 
and
\begin{equation}
  \label{M}
  M  \equiv  M_{pp'}=C_{Spp'} + C_{Npp'}, 
\end{equation}
with
\begin{eqnarray}
  C_{Spp'} & = & \sum_{\ell=0}^\infty\sum_{m=-\ell}^{\ell}
                  C_{\ell}B_{p\ell m}B^*_{p'\ell m} \times
                  \nonumber\\
              & & Y_{\ell m}({\bf x}_p)Y^*_{\ell m}({\bf x}_{p'}), 
                  \label{ClBB}\\
  C_{Npp'} & = & (A^T N^{-1}A)^{-1}.
  \label{CN}
\end{eqnarray}
Here $C_{Spp'}= \langle s_p s_{p'}^{T}\rangle$
is the pixel-pixel CMB signal correlation matrix,
and $C_{Npp'} = \langle n_p n_{p'}^{T}\rangle$
is the pixel-pixel noise correlation matrix.

We note first that in the estimation of $C_\ell$, 
although there exists  methods like the quadratic estimator 
(Bond et al.~1998)\markcite{BJK}
which avoid a direct evaluation of equation (\ref{L}),
the relationship between the beam expansion $B_{p\ell m}$
 and the power spectrum $C_\ell$ 
remains the same and is illustrated in equation (\ref{ClBB}).
Second, if the beam is identical for all pixels
and circularly symmetric, i.e., $B_{p\ell m}=B_{p'\ell m}=B_{\ell}$,
then equation (\ref{ClBB}) can be greatly simplified as
\begin{equation}
  \label{ClBB_sym}
  C_{Spp'} = \sum_{\ell=0}^\infty
                  \frac{2\ell+1}{4\pi}
                  C_{\ell}B^2_{\ell} P_\ell(\cos\theta_{pp'}),
\end{equation}
where $P_\ell$ is the Legendre function
and
$\theta_{pp'}=|{\bf x}_p-{\bf x}_{p'}|$ is the angular distance between the pixels.

Generally it is impractical to estimate  $C_\ell$ for all $\ell$
due to the constraints of finite sky coverage and computation power.
Instead, one divides the accessible $\ell$-range 
constrained by the sky coverage and the observing beam size
into several bands $\{b\}$,
and then estimates the band power $C_b$, 
i.e., one approximates $C_\ell$ in the form
\begin{equation}
  \label{ClCsh}
  C_\ell \approx C_b C_\ell^{\rm sh},
\end{equation}
where $C_\ell^{\rm sh}$ is a chosen shape function 
characterizing the scale dependence in each band.
For example, 
one can choose
\begin{equation}
  \label{Clsh}
  C_\ell^{\rm sh}=\frac{1}{\ell(\ell+1)},
\end{equation}
which leads to a scale-invariant form in each $\ell$ band,
i.e., $\ell(\ell+1)C_\ell={\rm const}\;\; \forall \ell \in b$.
With the approximation (\ref{ClCsh}),
one can rewrite equation (\ref{ClBB}) as
\begin{equation}
  \label{CBBB_app1}
  C_{Spp'} \approx
     \sum_{b}
     C_b
     {\cal K}_{pp'b}[C_\ell^{\rm sh},B_{p}, B_{p'}],
\end{equation}
where
\begin{eqnarray}
  {\cal K}_{pp'b}[C_\ell^{\rm sh},B_{p}, B_{p'}]
  =
   \sum_{\ell\in b}
      \sum_{m=-\ell}^{\ell}
         C_\ell^{\rm sh} \times
  \nonumber\\
         B_{p\ell m}B^*_{p'\ell m} 
          Y_{\ell m}({\bf x}_p)Y^*_{\ell m}({\bf x}_{p'}).
  \label{Kpp}
\end{eqnarray}
If the beam is symmetric,
then one has from equation (\ref{ClBB_sym}) or (\ref{CBBB_app1}) that 
\begin{equation}
  \label{CBBB_app2}
  C_{Spp'} \approx
     \sum_{b}
     {C}_b
     {\cal K}_b[\theta_{pp'};C_\ell^{\rm sh},B^2_\ell],
\end{equation}
where
\begin{eqnarray}
  &   {\cal K}_b[\theta_{pp'};C_\ell^{\rm sh},B^2_\ell]
        = \nonumber\\
  &                \sum_{\ell\in b}
                  \frac{2\ell+1}{4\pi}
                  C_\ell^{\rm sh}
                  B^2_\ell
                  P_\ell(\cos\theta_{pp'}).
  \label{K}
\end{eqnarray}

In the analysis procedure outlined above,
the first problem arises in equations (\ref{gamma_t}) and (\ref{s_p}).
Strictly speaking,
what is convolved in reality is not the pixel temperature in $s_p$ itself
but the CMB signal in the time-stream $\gamma_t$, i.e.,
\begin{equation}
  \label{gamma_t2}
  \gamma_t= 
  \sum_{\ell=0}^\infty\sum_{m=-\ell}^{\ell}
                   B_{t\ell m} a_{\ell m} Y_{\ell m}({\bf x}_t),
\end{equation}
where $B_{t\ell m}$ is the multipole expansion of 
the time-stream beam $B_t({\bf x})$.
 This means that the experiment gives us a beam which moves on the sky as
a function of time, $t$, and indeed may observe a different signal within the
same pixel, $p$, depending on the orientation of the beam and the
location of its center.  We thus make a map which may have many
different beams contributing to a single pixel.  However, in our
analysis formalism we must actually express this map as in (\ref{s_p}),
an observation of the sky with only a single pixel beam, $B_{p}$.  
Hence,
for the $C_\ell$ estimation,
we need to find a way to estimate the pixel-beam expansion $B_{p\ell m}$
from the $B_{t\ell m}$,
and
this will
be the focus of sections~\ref{the_effective_Bl_of_pixels}
and \ref{deconvolution_of_the_pixel_smoothing}.

The second problem appears in equation (\ref{CBBB_app1}).  If
the beam is not symmetric, the summation over $m$ and the dependence on
the pixel pair make the exact computation prohibitively expensive. 
To resolve this problem,
in section~\ref{the_effective_two_point_Bl} we introduce the 
pixel-pixel beam expansion $B_{\ell{\rm (eff)}}^2$,
which provides a consistent way to symmetrize asymmetric
beams.
This $B_{\ell{\rm (eff)}}^2$ then replaces
the $B_\ell^2$ in equation (\ref{CBBB_app2}),
so as to approximate equation (\ref{CBBB_app1}).

On general grounds,
the size of the observing beam is so small that,
when necessary,
we shall use the flat sky approximation
under the small-field limit.
This means that
when the size of a spherical patch is sufficiently small,
the expansion of the beam in spherical harmonics
is equivalent to 
a Fourier transform on a flat two-dimensional patch, i.e.,
\begin{equation}
  \label{sfl}
  B_{\ell m}
  = \int d\Omega B({\bf x})Y_{\ell m}({\bf x})
  \equiv 
    \int dx^2 B({\bf x}) e^{- i {\bf k} \cdot {\bf x}}
    = \widetilde{B}({\bf k}),
\end{equation}
and
\begin{equation}
  \ell \equiv k = |{\bf k}|.
\end{equation}
Throughout the paper,
we shall use a `tilde' to denote the Fourier transform of a quantity.

\section{THE CRITERIA FOR BEAM SYMMETRY}
\label{the_criteria_for_beam_symmetry}

It is important to clearly define the level of asymmetry of 
an antenna beam.
Consider the multipole expansion $B_{\ell m}$ of  the beam.
For a given $\ell$, the variance of $B_{\ell m}$ about its mean over $m$ is
\begin{equation}
  \eta_{\ell}^2=B_{\ell{\rm (ms)}}^2-B_{\ell{\rm (sm)}}^2,
\end{equation}
where $B_{\ell{\rm (ms)}}^2$ is the mean of squares over $m$:
\begin{equation}
  \label{B2lms}
  B_{\ell{\rm (ms)}}^2=\langle |B_{\ell m}|^2\rangle,
\end{equation}
and $B_{\ell{\rm (sm)}}^2$ is the square of the mean over $m$:
\begin{equation}
  \label{B2lsm}
  B_{\ell{\rm (sm)}}^2
  =\left[B_{\ell{\rm (m)}}\right]^2
  =\langle B_{\ell m}\rangle^2.
\end{equation}
Here $B_{\ell{\rm (m)}}$ is the mean of $B_{\ell m}$ over $m$,
and therefore can be either positive or negative.
We also note that
$B_{\ell{\rm (ms)}}^2$ is the power spectrum of the asymmetric beam,
and that
$B_{\ell{\rm (sm)}}^2$ is equivalent to 
the power spectrum of a symmetric beam
that is azimuthally averaged in the real space.
Based on this, one can define an `index of asymmetry' (IOA) as
\begin{equation}
  \label{varpi_l}
  \varpi_\ell=
     \frac{\eta_\ell}
     {B_{\ell{\rm (ms)}}}
     =
     \left[
     1-
     \frac{B_{\ell{\rm (sm)}}^2}
     {B_{\ell{\rm (ms)}}^2}
     \right]^{1/2}.
\end{equation}
We see that $\varpi_\ell$ varies from zero to one---the larger
the $\varpi_\ell$, the more asymmetric the beam.
We also note that
if the beam is symmetric, then $\varpi_\ell$ is exactly zero.
Thus for a given beam,
$\varpi_\ell$ provides us an objective measure
of the level of its asymmetry.

In certain situations,
we need to combine data from two or more photometers with different beam shapes.
We shall use a subscript $i$ ($i=0,1,2$, etc.) to denote the quantities
obtained from different photometers.
As an analog to equation (\ref{varpi_l}),
it proves useful to define an `index of combined asymmetry' (IOCA) for all the beams as
\begin{equation}
  \label{W_l}
  {\cal W}_\ell
     =
     \left[
     1-
     \frac{B_{\Sigma\ell{\rm (sm)}}^2}
    {B_{\Sigma\ell{\rm (ms)}}^2}
     \right]^{1/2},
\end{equation}
where 
\begin{eqnarray}
  B_{\Sigma\ell{\rm (sm)}}^2 & = &
 \left[
  \sum_i \zeta_i B_{i\ell{\rm (m)}}
 \right]^{2}, 
 \label{BIsm2}\\
  B_{\Sigma\ell{\rm (ms)}}^2 & = &
 \left[
  \sum_i \zeta_i \sqrt{B_{i\ell{\rm (ms)}}^2}
 \right]^{2},
 \label{BIms2}
\end{eqnarray}
the $B_{i\ell{\rm (m)}}$ and $B_{i\ell{\rm (ms)}}^2$
are the $B_{\ell{\rm (m)}}$ and $B_{\ell{\rm (ms)}}^2$
of photometer $i$ respectively,
\begin{equation}
  \label{zeta}
  \zeta_i = \frac{{t_{i\rm (obs)}}/{{\rm NET}_i^2}}
      {\sum_i \left[{t_{i\rm (obs)}}/{{\rm NET}_i^2}\right]},
\end{equation}
$t_{i\rm (obs)}$ is the total observation time of photometer $i$,
and ${\rm NET}_i$ is its noise equivalent temperature (NET).
Here $B_{\Sigma\ell{\rm (sm)}}^2$ is the square of the noise-weighted mean of $B_i$,
and $B_{\Sigma\ell{\rm (ms)}}^2$ is the noise-weighted mean of the squares of $B_i$
assuming all the $B_i$ are fully correlated.
As one can see,
the ${\cal W}_\ell$ varies between zero and one---the 
larger the ${\cal W}_\ell$, 
the more asymmetric a $\zeta_i$-weighted combined beam can be
(depending on the detailed orientations of the beams
in the temporal samples; 
we shall discuss this later).
This also means that
the IOA ($\varpi_\ell$) of an average beam with a weight $\zeta_i$
for each $B_i$
is always equal to or smaller than ${\cal W}_\ell$,
although the individual $\varpi_{i\ell}$ of $B_i$
may be larger than ${\cal W}_\ell$.
If ${\cal W}_\ell=0$,
then we know that
all the beams are symmetric ($\varpi_{i\ell}=0$), and vice versa.

For the purpose of power spectrum estimation,
one can employ $\varpi_{i\ell}$
(or ${\cal W}_\ell$ when combining data of different observing beams)
to decide if a simple symmetric-beam approximation is sufficient.
For example,
at $\ell$'s where $\varpi_\ell \approx 0$,
we expect equation (\ref{CBBB_app2}) to be adequate.
On the other hand,
at $\ell$'s where $\varpi_\ell$ (or ${\cal W}_\ell$) deviates significantly from zero,
one may need to employ equation (\ref{CBBB_app1}).
We shall further discuss these situations,
and
the use of the IOA ($\varpi_\ell$)
and the IOCA (${\cal W}_\ell$) later.

\section{THE AVERAGE PIXEL-BEAM EXPANSION} 
\label{the_effective_Bl_of_pixels}

\subsection{The pixel-beam expansion}
\label{the_pixel_beam_expansion}

We first estimate the `pixel-beam expansion', $B_{p\ell m}$, 
from given observing beams, $B_{t\ell m}$.
A naive way to investigate this
is to substitute equations (\ref{s_p}) and (\ref{gamma_t2})
into the model (\ref{gamma_t}), leading to
\begin{equation}
\label{eq:xpxt}
  A_{tp} B_{p\ell m}  Y_{\ell m}({\bf x}_p) =  
  B_{t\ell m}  Y_{\ell m}({\bf x}_t).
\end{equation}
This equation holds if and only if 
there exists one ${\bf x}_p$ for every ${\bf x}_t$ such that
${\bf x}_t={\bf x}_p$.
In this case, we have $B_{p\ell m}=B_{t\ell m}$.
This is of course true when the pixel size is infinitesimal,
but is unlikely to be fulfilled in reality.
Nevertheless,
equation (\ref{eq:xpxt}) is just the result of the modeling
and therefore not necessarily a requirement in practice.
In our formalism for the power spectrum estimation,
the $s_p$ is an unknown quantity to be estimated by
using equation (\ref{m}),
so 
the actual relation
between $B_{p\ell m}$ and $B_{t\ell m}$ 
should be also obtained through the same process.
First,
we substitute equation (\ref{d_t}) into (\ref{m}), 
and the CMB signal part yields
\begin{equation}
  \label{sp}
  s_p = C_N A^T N^{-1} \gamma_t,
\end{equation}
where $C_N\equiv C_{Npp'}$ as defined in equation (\ref{CN}).
Further substituting equations (\ref{s_p}) and (\ref{gamma_t2})
into this result,
we obtain
\begin{equation}
  \label{BplmY0}
  B_{p\ell m} Y_{\ell m}({\bf x}_p) = C_N A^T N^{-1} B_{t \ell m}
  Y_{\ell m}({\bf x}_t).
\end{equation}
This equation is completely general, 
and should be in principle satisfied
when one tries to find the $B_{p\ell m}$ from the given $B_{t\ell m}$.
We thus see that
equation (\ref{eq:xpxt}) is just
one of the solutions to equation (\ref{BplmY0}),
but not necessarily a requirement
for the purpose of power spectrum estimation.

In most cases,
the noise $n_t$ in each temporal measure is nearly independent
from the others,
so
the time-time noise correlation matrix $N_{tt'}$ is diagonal,
with the $tt$ elements equal to the noise variance
 at each time sample,
i.e.,
\begin{equation}
  \label{white-noise}
  N_{tt'} = \mu_t^2\delta(t-t'),
\end{equation}
where  $\mu_t$ is the standard deviation of time sample $t$,
and $\delta(t-t')$ is a Dirac Delta. 
This allows us to simplify equation (\ref{BplmY0}) as
\begin{equation}
  \label{BplmY}
  B_{p\ell m} Y_{\ell m}({\bf x}_p)
  = \sum_{t\in p}\xi_t B_{t\ell m} Y_{\ell m}({\bf x}_t),
\end{equation}
where
$\xi_t$ is the noise-estimated statistical weight at $t$:
\begin{equation}
  \label{xi_t}
  \xi_t=\frac{\mu_t^{-2}}{\sum_{t\in p} \mu_t^{-2}}.
\end{equation}
For simplicity,
we shall take this white-noise assumption for further investigation.
We consider the more general case of correlated noise in the Appendix, 
and show that
this white-noise approximation is appropriate in most practical cases.
The conditions for the use of this white-noise assumption
will be also derived in the Appendix (see eq.~[\ref{white-noise-cond}]).

To further simplify equation (\ref{BplmY}),
we assume that ${\bf x}_t \equiv {\bf x}_p$ $\forall t\in p$
(i.e., the temporal measure $\gamma_t$ is thought of as a `sample'
of the pixel temperature $s_p$; see eq.~[\ref{sp}]),
so that the pixel-beam expansion can now be obtained as
\begin{equation}
  \label{Bplm}
  B_{p\ell m} 
  = \sum_{t\in p}\xi_t B_{t\ell m}.
\end{equation}
The assumption, ${\bf x}_t \equiv {\bf x}_p$,
for achieving this result will be
relaxed in section~\ref{deconvolution_of_the_pixel_smoothing}, where we
show that only an extra correction is required.

\subsection{The average pixel-beam expansion}

As will be shown,
it proves useful to remove the pixel dependence of $B_{p\ell m}$
in the formalism of the $C_\ell$ estimation.
We thus consider the noise-weighted average of $B_{p\ell m}$ over all pixels
(c.f.\ eq.~[\ref{m}]):
\begin{equation}
  \label{barBplm00}
  \overline{B}_{p\ell m}
   =  H (U^T C_N^{-1} B_{p\ell m}),
\end{equation} 
where $U\equiv U_p$ is a contraction vector 
with entries all equal to unity, and
\begin{equation}
  \label{H}
  H = (U^T C_N^{-1} U)^{-1} .
\end{equation}
We shall call $\overline{B}_{p\ell m}$
the `average pixel-beam expansion'.
We note that
the subscript $p$ in $\overline{B}_{p\ell m}$
does not mean the pixel dependence as in the usual convention,
but
indicates that this is a mean taken over all pixels.

With the white-noise assumption (eq.~[\ref{white-noise}]),
the $\overline{B}_{p\ell m}$ can be calculated explicitly 
by substituting equation (\ref{Bplm}) into equation (\ref{barBplm00}):
\begin{equation}
  \label{barBplm2}
  \overline{B}_{p\ell m}
  = \sum_t \chi_t B_{t\ell m},
\end{equation}
where
\begin{equation}
  \label{chi_t}
  \chi_t = \frac{\mu_t^{-2}}{\sum_{t} \mu_t^{-2}}.
\end{equation}
If the data are from a single photometer with a constant noise level,
then equation (\ref{barBplm2}) reduces to a simple linear average
of all time-stream beams.
If the data are combined from different photometers,
then the $\mu_t$ can be approximated as
(c.f.\ eq.~[\ref{zeta}])
\begin{equation}
  \label{mu_t}
  \mu_t=
   \frac{{\rm NET}_t}{\sqrt{\delta t_{\rm (obs)}}},
\end{equation}
where
${\rm NET}_t$ is the NET of
the corresponding photometer at time $t$,
and $\delta t_{\rm (obs)}$ is the integration time of the
temporal observation at $t$.
If the integration time remains unchanged among photometers,
then the $\mu_t$ in equation (\ref{chi_t})
can be simply taken as the NET of the corresponding photometer.
We also note that
with the definition (\ref{mu_t}),
equations (\ref{zeta}) and (\ref{chi_t}) can be related as
\begin{equation}
  \label{zeta_chi}
  \zeta_i=\sum_{t\in i} \chi_t,
\end{equation}
meaning that
$\zeta_i$ is the total noise-estimated weight of photometer $i$.

We note that 
in cases where 
both
the shape of the experimental beam
and its orientation relative to the pixel
are roughly constant throughout the observation,
we have a reasonable approximation (see eq.~[\ref{barBplm2}]):
\begin{equation}
  \label{barBplmsame}
  \overline{B}_{p\ell m} \approx B_{t\ell m}.
\end{equation}
In other cases,
equation (\ref{barBplm2}) will need to be employed,
for example, when the relative orientation between the asymmetric beam
and the pixels changes, 
or when data from different photometers
 are combined together.
We also note that
even if all the beams $B_i$ of different photometers
are symmetric (i.e., $\varpi_{i\ell}={\cal W}_\ell=0$),
the ${B}_{p\ell m}$ may still have pixel dependence
due to the various relative contribution of $B_i$ within different pixels
(see eq.~[\ref{Bplm}]).
In such cases,
one will need to consider equation (\ref{barBplm2}),
and
a simple formalism like equation (\ref{CBBB_app2}) will be invalid 
for the estimation of the CMB angular power spectrum,
since the ${B}_{p\ell}^2$ is different on each pixel.
As will be shown,
the formalism we shall develop is also capable of dealing with
this situation.

\subsection{Useful Limits}
\label{pbexp_limit}

We now derive useful constraints
on the magnitude of the average pixel-beam expansion  $\overline{B}_{p\ell m}$.
In the small-field limit,
the power spectrum of $\overline{B}_{p\ell m}$ can be written as
(see eqs.~[\ref{sfl}], [\ref{B2lms}], and [\ref{barBplm2}])
\begin{equation}
  \label{Bl2eff_2}
  \overline{B}^2_{p\ell{\rm (ms)}}
  \equiv
  \frac{1}{\pi}
    \int_0^\pi d\varphi
    \left|\sum_{t} \chi_t \widetilde{B}_{t}({\bf k})
    \right|^{2},
\end{equation}
where
$\varphi$ is the phase angle of ${\bf k}$ on the ring $|{\bf k}|=k$.
We first consider single-photometer experiments.
In this case,
if the beam pattern remains the same throughout the entire observation
but with only different orientations at different $t$,
then we can rewrite $B_t$ as
\begin{equation}
  \label{B0}
   B_t=A(\beta_t) B_0,
\end{equation}
where $A(\beta_t)$ is the rotation matrix,
$\beta_t$ is the rotation angle at time $t$ with respect to $t=0$,
and $B_0$ is the shape of the time-stream beam at $t=0$.
Substituting this into equation (\ref{Bl2eff_2}) gives
\begin{equation}
  \label{Bl2eff_3}
  \overline{B}^2_{p\ell{\rm (ms)}}
  \equiv
  \frac{1}{\pi}
    \int_0^\pi d\varphi
    \left|
      \int_{0}^{2\pi} f(\beta) A(\beta)\widetilde{B}_0({\bf k})d\beta
    \right|^{2},
\end{equation}
where $f(\beta)$ is the weighting function of a rotation angle $\beta$,
and satisfies $\int_0^{2\pi}d\beta f(\beta)=1$.  It is then
straightforward to show that the function $f(\beta)$ that minimizes the
right hand side of the above equation is $f(\beta)=1/2\pi$, leading to
\begin{equation}
  \label{Bl2eff_case1}
  \left.
  \overline{B}^2_{p\ell{\rm (ms)}}
  \right|_{\rm min}
  \equiv
  \left[
    \frac{1}{\pi}
    \int_{0}^{\pi} d\varphi
      \widetilde{B}_0({\bf k})
  \right]^2
  \equiv
  B_{0\ell{\rm (sm)}}^2,
\end{equation}
where $B_{0\ell{\rm (sm)}}^2$ is as defined in equation (\ref{B2lsm}).
On the other hand,
the function $f(\beta)$ that maximizes the right hand side
of equation (\ref{Bl2eff_3})
is $f(\beta)=\delta(\beta-\beta_0)$ 
(Dirac Delta, $\beta_0 \in \{0, 2\pi\}$),
and this gives
\begin{equation}
  \label{Bl2eff_case2}
  \left.
  \overline{B}^2_{p\ell{\rm (ms)}}
  \right|_{\rm max}
  \equiv
    \frac{1}{\pi}
    \int_{0}^{\pi} d\varphi
  \left|
    \widetilde{B}_0({\bf k})
  \right|^2
  \equiv
  B_{0\ell{\rm (ms)}}^2,
\end{equation}
where $B_{0\ell{\rm (ms)}}^2$ is as defined in equation (\ref{B2lms}).
These results tell us that
when the pixels are scanned almost uniformly in all directions,
then the resulting $\overline{B}^2_{p\ell{\rm (ms)}}$
should be closer to 
$\left.
  \overline{B}^2_{p\ell{\rm (ms)}}
  \right|_{\rm min} =
B_{0\ell{\rm (sm)}}^2$.
When the pixels are scanned with an almost fixed direction,
then the resulting $\overline{B}^2_{p\ell{\rm (ms)}}$ should be closer to 
$\left.
  \overline{B}^2_{p\ell{\rm (ms)}}
  \right|_{\rm max} =
B_{0\ell{\rm (ms)}}^2$.
Thus,
we have a good check of the numerically calculated
$\overline{B}_{p\ell m}$
from equation (\ref{barBplm2}) (or eq.~[\ref{Bl2eff_3}]),
i.e.,
a constraint on
the amplitude of $\overline{B}_{p\ell{\rm (ms)}}^2$:
\begin{equation}
  \label{B2l_constraint}
  B_{0\ell{\rm (ms)}}^2 
  \geq \overline{B}_{p\ell{\rm (ms)}}^2 
  \geq B_{0\ell{\rm (sm)}}^2,
\end{equation}
or equivalently,
\begin{equation}
  \label{B2l_constraint2}
  1 \geq \frac{\overline{B}_{p\ell{\rm (ms)}}^2}{B_{0\ell{\rm (ms)}}^2} 
  \geq 1-\varpi_{0\ell}^2,
\end{equation}
where $\varpi_{0\ell}$ is the IOA of $B_0$.
For symmetric beams,
all the equality signs hold.
In experiments,
one can take $B_0$ as the measured beam,
and then use equation (\ref{varpi_l}) to calculate $\varpi_{0\ell}$.

When we combine data from two or more photometers
with different beam shapes,
following the same line of development as above gives 
(see eqs.~[\ref{BIsm2}], [\ref{BIms2}], [\ref{barBplm2}], and [\ref{zeta_chi}])
\begin{equation}
  \label{B2l2_constraint}
  B_{\Sigma\ell{\rm (ms)}}^2
  \geq \overline{B}_{p\ell{\rm (ms)}}^2
  \geq B_{\Sigma\ell{\rm (sm)}}^2,
\end{equation}
or equivalently,
\begin{equation}
  \label{B2l2_constraint2}
  1 \geq \frac{\overline{B}_{p\ell{\rm (ms)}}^2}
      {B_{\Sigma\ell{\rm (ms)}}^2} 
  \geq 1-{\cal W}_{\ell}^2,
\end{equation}
where ${\cal W}_{\ell}$ is the IOCA defined in equation (\ref{W_l}).
We shall further discuss the use of these limits later.

\section{THE PIXEL-PIXEL BEAM EXPANSION}
\label{the_effective_two_point_Bl}

\subsection{Formalism}

In the data analysis procedure 
briefly demonstrated in section~\ref{the_convention},
the effect of asymmetric beam convolution
manifests itself in equation (\ref{ClBB}).
However,
the summation over $m$ and the dependence on the pixel pair
make it computationally expensive.
Therefore,
we prefer to use the form of
equation (\ref{ClBB_sym}) as an approximation.
This can be achieved by replacing the $B_\ell^2$ in equation (\ref{ClBB_sym})
with a `pixel-pixel beam expansion' $B_{\ell{\rm (eff)}}^2$,
which we shall derive in this section.

First,
one can replace the $B_\ell^2$ in equation (\ref{ClBB_sym}) with
\begin{equation}
  \label{Bl2pp}
  {B}^2_{pp'\ell}=
      \frac{{4\pi}\sum_{m}
                  B_{p\ell m}B^*_{p'\ell m}
                  Y_{\ell m}({\bf x}_p)Y^*_{\ell m}({\bf x}_{p'})}
        {(2\ell+1) P_\ell(\cos\theta_{pp'})},
\end{equation}
so that equation (\ref{ClBB_sym}) is equivalent to equation (\ref{ClBB}).
In the small-field limit,
equation (\ref{Bl2pp}) becomes
\begin{equation}
  \label{Bl2pp_sfa}
  {B}^2_{pp'\ell} \equiv
  {B}^2_{pp'k} =
   \frac{
        {\cal J}\left[
           k\Delta x,\varphi_0; {\widetilde{\cal B}}^2_{pp'}({\bf k})
         \right]
        }{
        J_0( k \Delta x)
        },
\end{equation}
where
\begin{eqnarray}
 & {\cal J}\left[k\Delta x,\varphi_0; {\widetilde{\cal B}}^2_{pp'}({\bf k})\right]
  =  \nonumber\\
 & \frac{1}{\pi}
        \int_0^\pi d\varphi
           \left\{
               \Re\left[{\widetilde{\cal B}}^2_{pp'}({\bf k})\right] 
               \cos[k\Delta x \cos(\varphi-\varphi_0)]
           \right.
    \nonumber\\
 &  -
           \left.
               \Im\left[{\widetilde{\cal B}}^2_{pp'}({\bf k})\right]
               \sin[k\Delta x \cos(\varphi-\varphi_0)]
           \right\},
  \label{J}
\end{eqnarray}
 $\Delta{\bf x}={\bf x}_p-{\bf x}_{p'}$, 
$\Delta{x}=|\Delta{\bf x}|\equiv\theta_{pp'}$,
$\varphi_0$ is the phase angle of $\Delta{\bf x}$,
$J_0$ is the Bessel function of the first kind of integral order $0$,
${\widetilde{\cal B}}^2_{pp'}({\bf k})=
\widetilde{B}_{p}({\bf k})\widetilde{B}^*_{p'}({\bf k})$,
and
$\Re$ and $\Im$ indicate the real and imaginary parts of
${\widetilde{\cal B}}^2_{pp'}$ respectively.
We notice that ${\cal J}[k\Delta x,\varphi_0; 1]=J_0( k \Delta x)$.
Therefore if the beam is circularly symmetric
and remains the same on all pixels,
i.e., ${\widetilde{\cal B}}^2_{pp'}({\bf k})\equiv \widetilde{B}^2_k$,
then 
${\cal J}[k\Delta x,\varphi_0; {\widetilde{\cal B}}^2_{pp'}({\bf k})]
={\cal J}[k\Delta x,\varphi_0; \widetilde{B}^2_k]
=J_0( k \Delta x) \widetilde{B}^2_k$,
so that 
${B}^2_{pp'k}$ in equation (\ref{Bl2pp_sfa})
becomes $\widetilde{B}^2_k$ exactly
as required.

To save computation time and memory when estimating $C_l$,
we need to remove the dependence of ${B}^2_{pp'\ell}$
on the particular choice of a pixel pair $({\bf x}_p,{\bf x}_{p'})$.
We achieve this by taking the average of ${B}^2_{pp'\ell}$
over all possible $({\bf x}_p,{\bf x}_{p'})$ pairs:
\begin{equation}
  \label{Bl2eff_avg}
  {B}^2_{\ell{\rm (eff)}}=
    \left\langle
      {B}^2_{pp'\ell}
    \right\rangle.
\end{equation}
We call this ${B}^2_{\ell {\rm (eff)}}$
the `pixel-pixel beam expansion'.

Even with this,
equation (\ref{Bl2eff_avg}) together with equation (\ref{Bl2pp_sfa})
is still computationally expensive and may not be feasible.
Therefore 
we further simplify the formalism in the following way.
First,
we remove the dependence of 
${\widetilde{\cal B}}^2_{pp'}({\bf k})$ in equation (\ref{Bl2pp_sfa})
on pixel pairs, 
by replacing it with a noise-weighted average (c.f.\ eqs.~[\ref{m}] and [\ref{barBplm00}])
\begin{equation}
  \label{B2ppk}
  \overline{\widetilde{\cal B}_{pp'}^2}
  =
  (U^T C_N^{-1} UU^T C_N^{-1}U)^{-1}
  (U^T C_N^{-1} \widetilde{\cal B}_{pp'}^2 C_N^{-1}U).
\end{equation}
Here the subscript $pp'$ in $\overline{\widetilde{\cal B}_{pp'}^2}$ does not
mean the pixel pair dependence as in the usual convention,
but indicates that
the mean is taken over all pixel pairs.
With this replacement,
equation (\ref{Bl2pp_sfa}) is now only a function of $\Delta {\bf x}$
for a given $\ell\equiv k$.
Thus when evaluating equation (\ref{Bl2eff_avg}),
we can classify all possible $\Delta{\bf x}$ into
several groups of different $\Delta x$,
each with several subgroups of different $\varphi_0$.
This gives
\begin{equation}
  \label{Bl2eff_avg2}
  {B}^2_{\ell{\rm (eff)}}
  \equiv
    \sum_{\Delta x, \varphi_0}
        g(\Delta x, \varphi_0)
      \frac{
          {\cal J}\left[k\Delta x,\varphi_0; 
            \overline{\widetilde{\cal B}_{pp'}^2}({\bf k})\right]
          }{
          J_0( k \Delta x)
          },
\end{equation}
where $g(\Delta x, \varphi_0)$ is
the weight of the configuration $(\Delta x, \varphi_0)$,
i.e.,
the number of pixel pairs with $\Delta x$ and $\varphi_0$,
divided by the total number of pixel pairs.
It satisfies $\sum_{\Delta x, \varphi_0}g(\Delta x, \varphi_0)=1$.
This algorithm can normally reduce the number of operations in
equation (\ref{Bl2eff_avg}) by several orders of  magnitude,
because the element number of $\{(\Delta x, \varphi_0)\}$
is normally several orders below that of $\{({\bf x}_p,{\bf x}_{p'})\}$.
In addition,
if the number of pixels is large enough as in most cases,
then $\varphi_0$ is nearly uniformly distributed between $0$ and $2\pi$
for every given $\Delta x$,
depending on the relative locations of all pixels.
In this case,
after the summation over $\varphi_0$ at each given $\Delta x$
in equation (\ref{Bl2eff_avg2}),
the first term inside the integral in equation (\ref{J})
(which enters eq.~[\ref{Bl2eff_avg2}])
becomes 
$\Re\left[\overline{\widetilde{\cal B}_{pp'}^2}({\bf k})\right]J_0( k \Delta x)$
and the second term vanishes.
Thus
the Bessel function in equation (\ref{Bl2eff_avg2})
can be removed 
and we have 
\begin{equation}
  \label{Bl2eff_avg3}
  {B}^2_{\ell{\rm (eff)}}
  \approx
  \frac{1}{\pi}
        \int_0^\pi 
          \Re\left[\overline{\widetilde{\cal B}_{pp'}^2}({\bf k})\right]
        d\varphi.
\end{equation}
With careful simplification of 
the real part of equation (\ref{B2ppk}), 
we also find that
\begin{equation}
  \label{B2ppk2}
  \Re\left[\overline{\widetilde{\cal B}_{pp'}^2}({\bf k})\right]
  =
  \left|\overline{\widetilde{B}}_p({\bf k})\right|^2,
\end{equation}
where $\overline{\widetilde{B}}_p({\bf k})\equiv \overline{B}_{p\ell m}$ 
as defined in equation (\ref{barBplm00}).
We note that
the average over all pixel pairs
(the left-hand side of eq.~[\ref{B2ppk2}])
is now reduced to
the average  over all pixels (the right-hand side).
This further enables us to simplify equation (\ref{Bl2eff_avg3}) as
\begin{equation}
  \label{Bl2eff_avg4}
  {B}^2_{\ell{\rm (eff)}}
  \approx
  \frac{1}{\pi}
        \int_0^\pi 
          \left|\overline{\widetilde{B}}_p({\bf k})\right|^2
        d\varphi
  \equiv
  \overline{B}^2_{p\ell{\rm (ms)}},
\end{equation}
where the last step uses the definition (\ref{B2lms}),
and the $\overline{B}^2_{p\ell{\rm (ms)}}$ is readily
evaluated in equation (\ref{Bl2eff_2}).
When calculating $\overline{B}^2_{p\ell{\rm (ms)}}$,
one can take the form of equation (\ref{Bl2eff_3})
to save computation time.
We note that
the approximation sign above will become equality
when $\varphi_0$ is uniformly distributed between $0$ and $2\pi$.
In section~\ref{numerical_verifications},
we shall numerically verify this result.


With such,
now we can use the form of equation (\ref{ClBB_sym}) 
to approximate equation (\ref{ClBB})
in the presence of asymmetric beams
or when combining data with different symmetric beams.
In other words,
we have equation (\ref{ClBB}) being approximated as
\begin{equation}
  \label{ClBB_app}
  C_{Spp'} \approx
                  \sum_{\ell=0}^\infty
                  \frac{2\ell+1}{4\pi}
                  {C}_{\ell}
                  {B}^2_{\ell{\rm (eff)}}
                  P_\ell(\cos\theta_{pp'}).
\end{equation}
Furthermore,
as illustrated in equations (\ref{ClCsh}) through (\ref{K}) 
and the context,
one normally divides the $\ell$ range under investigation into several bands,
due to the finite sizes of the sky coverage and the observing beam,
as well as the limited computation power.
Using this formalism,
we can approximate equation (\ref{CBBB_app1})
using equation (\ref{CBBB_app2}) with its $B_\ell^2$ replaced
by the ${B}^2_{\ell{\rm (eff)}}$ calculated above.
This gives
\begin{equation}
  \label{CBBB_app3}
  C_{Spp'} \approx
     \sum_{b}
     {C}_b
     {\cal K}_b[\theta_{pp'};C_\ell^{\rm sh},B^2_{\ell{\rm (eff)}}].
\end{equation}

\subsection{Uncertainties}

When making the approximation (\ref{ClBB_app}),
we inevitably induce errors in the basis 
${B}^2_{\ell{\rm (eff)}}
P_\ell(\cos\theta_{pp'})$ for each pixel pair.
These errors can be represented as
\begin{equation}
  \label{Bl2eff}
  \frac{
    B^2_{pp'\ell}
  }{
    {B}^2_{\ell{\rm (eff)}}
  }
  \equiv
  1 \pm \sigma_{\ell},
\end{equation}
where
$\sigma_{\ell}$ is the normalized standard deviation of the errors.
This deviation can be simultaneously evaluated
while one performs equation (\ref{Bl2eff_avg2}), i.e.,
\begin{eqnarray}
  \sigma_{\ell}^2
  = 
  \sum_{\Delta x, \varphi_0}
  g(\Delta x, \varphi_0)
  \left\{
    \frac{
        {\cal J}\left[k\Delta x,\varphi_0; 
            \overline{\widetilde{\cal B}_{pp'}^2}\right]
        }{
        {B}^2_{\ell{\rm (eff)}}
        J_0( k \Delta x)
        }
  \right\}^2   - 1.
  \label{sigmal2}
\end{eqnarray}
Since $C_\ell$ appears in combination with
${B}^2_{\ell{\rm (eff)}}
P_\ell(\cos\theta_{pp'})$ (see eq.~[\ref{ClBB_app}]),
we know that 
$\sigma_{\ell}$ basically quantifies the bias in $C_\ell$
for each individual pixel pair.
Nevertheless,
the resulting bias in the final $C_\ell$ estimates
by using the approximation (\ref{ClBB_app})
together with the likelihood analysis (see eq.~[\ref{L}] and context)
may be much smaller than $\sigma_{\ell}$,
because the resulting $C_\ell$ is a consequence of
the contribution from {\em all} pixel pairs.
For example,
if all pixel pairs contribute to the likelihood function (\ref{L})
as a linear combination of 
${B}^2_{pp'\ell} P_\ell(\cos\theta_{pp'})$,
then the resulting bias in $C_\ell$ will be as small as
$\sigma_{\ell}/{\cal N}_p$,
where ${\cal N}_p$ is the total  number of pixels.
Although we know that reality is not like such a simple case,
we can still quantify the bias of approximation (\ref{ClBB_app})
using numerical simulations.

Similarly,
we can consider the errors in the band power ${C}_b$
for each individual pixel pair,
resulting from the approximation (\ref{CBBB_app3}).
Since $C_b$ is coupled with ${\cal K}_b$ (eq.~[\ref{CBBB_app3}])
or ${\cal K}_{pp'b}$ (eq.~[\ref{CBBB_app1}]),
the errors in $C_b$ for each individual pixel pair
may be quantified 
by comparing ${\cal K}_b$ and ${\cal K}_{pp'b}$,
as we did for ${B}^2_{\ell{\rm (eff)}}$ and ${B}^2_{pp'\ell}$.
However,
as argued earlier,
the result calculated in this way
quantifies only the errors in $C_b$ for each individual pixel pair,
and
the real bias of the approximation (\ref{CBBB_app3})
together with the likelihood analysis
may be much smaller.
We shall quantify the real systematic bias of this approximation
in section~\ref{numerical_verifications},
using numerical simulations.

\section{SYMMETRY VS.\ ASYMMETRY}
\label{symmetry_vs_asymmetry}

In this section,
we investigate the conditions 
under which one needs to employ the formalism for treating asymmetric beams,
i.e.\ the formalism we developed in the previous two sections.
We first consider the case
where the data to be analyzed
is from only one photometer.
From equation (\ref{Bl2eff_avg4}),
we know that
the pixel-pixel beam expansion $B_{\ell{\rm (eff)}}^2$
can be approximated by $\overline{B}^2_{p\ell{\rm (ms)}}$,
and therefore should be also constrained by equation (\ref{B2l_constraint2}), 
resulting in
\begin{equation}
  \label{B2leff_constraint}
  1 \geq \frac{B_{\ell{\rm (eff)}}^2}{B_{0\ell{\rm (ms)}}^2} 
  \geq 1-\varpi_{0\ell}^2.
\end{equation}
This implies that
if we simply use $B_{0\ell{\rm (ms)}}^2$
(where $B_0$ is the measured beam shape from the experiment)
as the $B_{\ell{\rm (eff)}}^2$ in our formalism,
then $B_{\ell{\rm (eff)}}^2$ will be overestimated
by at most $\varpi_{0\ell}^2/(1-\varpi_{0\ell}^2)$.

Furthermore, 
we consider the errors in $C_\ell$ resulting from this effect.
In our formalism,
the beam convolution appears as the multiplication of
$B_{\ell{\rm (eff)}}^2$ and $C_\ell$
(see eq.~[\ref{ClBB_app}]),
so the errors in $C_\ell$ can be expressed as
\begin{equation}
  \label{delta_Cl}
  \delta_{C_\ell}
    =  \frac{dC_\ell}{C_\ell}
    =  \frac{dB_{\ell{\rm (eff)}}^{-2}}{B_{\ell{\rm (eff)}}^{-2}}.
\end{equation}
Taking 
$dB_{\ell{\rm (eff)}}^{-2}
=B_{0\ell{\rm (ms)}}^{-2}-B_{\ell{\rm (eff)}}^{-2}$,
we have from equations (\ref{B2leff_constraint}) and (\ref{delta_Cl}) that 
\begin{equation}
  \label{Cl_constraint}
  0 \geq \delta_{C_\ell}
  \geq -\varpi_{0\ell}^2.
\end{equation}
This means that
when we use $B_{0\ell{\rm (ms)}}^2$
as the $B_{\ell{\rm (eff)}}^2$ in our formalism,
then the resulting $C_\ell$ at a given $\ell$
will be {\em underestimated} by at most $\varpi_{0\ell}^2$.
To share this error on both sides of a mis-estimated $C_\ell$,
we can choose the $B_{\ell{\rm (eff)}}^2$ to be
\begin{equation}
  \label{B20lmid}
  {B_{0\ell{\rm (mid)}}^2}
  =
  \frac{2}{B_{0\ell{\rm (ms)}}^{-2}+B_{0\ell{\rm (sm)}}^{-2}}
  =
  \frac{1-\varpi_{0\ell}^2}{1-\varpi_{0\ell}^2/2} B_{0\ell{\rm (ms)}}^{2},
\end{equation}
so that the resulting error in the $C_\ell$ estimates is now constrained as
\begin{equation}
  \label{Clmid_constraint}
  \frac{\varpi_{0\ell}^2}{2-2\varpi_{0\ell}^2}
  \geq \delta_{C_\ell{\rm (mid)}}
  \geq -\frac{\varpi_{0\ell}^2}{2}.
\end{equation}
When $\varpi_{0\ell}^2\ll 1$,
we have $|\delta_{C_\ell{\rm (mid)}}|  \lesssim \varpi_{0\ell}^2/2$.
If the beam is symmetric,
then all the equality signs above hold and 
$B_{0\ell{\rm (ms)}}^2 = B_{0\ell{\rm (mid)}}^2 = 
B_{0\ell{\rm (sm)}}^2 = B_{\ell{\rm (eff)}}^2$.

Following the same line of logic,
we now consider the cases where
the data to be analyzed is combined from two or more photometers.
In this case,
it is also straightforward to show that
if we choose the $B_{\ell{\rm (eff)}}^2$ to be
(see also eqs.~[\ref{BIsm2}] and [\ref{BIms2}] for definitions)
\begin{equation}
  \label{B2ilmid}
  {B_{\Sigma\ell{\rm (mid)}}^2}
  =  \frac{2}{B_{\Sigma\ell{\rm (ms)}}^{-2}+B_{\Sigma\ell{\rm (sm)}}^{-2}}
  =  \frac{1-{\cal W}_{\ell}^2}{1-{\cal W}_{\ell}^2/2}
  B_{\Sigma\ell{\rm (ms)}}^{2},
\end{equation}
then the errors in the $C_\ell$ estimates are constrained as
\begin{equation}
  \label{Cl2mid_constraint}
  \frac{{\cal W}_{\ell}^2}{2-2{\cal W}_{\ell}^2}
  \geq \delta_{C_\ell{\rm (mid)}}
  \geq -\frac{{\cal W}_{\ell}^2}{2}.
\end{equation}
When ${\cal W}_{\ell}^2\ll 1$,
we have $|\delta_{C_\ell{\rm (mid)}}|  \lesssim {\cal W}_{\ell}^2/2$.
If all the beams are symmetric ($\varpi_{i\ell}=0$),
then all the equality signs above hold.

As a result,
we see that
if the $\varpi_{0\ell}^2/2$ (or ${\cal W}_{\ell}^2/2$)
is well below the tolerated maximum error of $C_\ell$,
then we can use $B_{0\ell{\rm (mid)}}^2$ (or $B_{\Sigma\ell{\rm (mid)}}^2$)
as the $B_{\ell{\rm (eff)}}^2$
in the symmetric-beam formalism,
i.e.,
we can simply use equation (\ref{CBBB_app2})
with $B_{\ell}^2 = B_{0\ell{\rm (mid)}}^2$ (or $B_{\Sigma\ell{\rm (mid)}}^2$),
without the need of going through the procedure developed 
in sections~\ref{the_effective_Bl_of_pixels}
and \ref{the_effective_two_point_Bl}.
The associated errors in the final $C_\ell$ estimates will be constrained
by equation (\ref{Clmid_constraint}) (or eq.~[\ref{Cl2mid_constraint}]).

\section{UNCERTAINTIES FROM BEAM MEASUREMENT}
\label{uncertainties_from_beam_measurement}

It is inevitable for any experiment that
there are uncertainties in the measurement of the beam shapes.
It is therefore crucial to quantify the uncertainties in the final
$C_\ell$ estimates resulting from this beam shape uncertainties.
For a given beam $B({\bf x})$,
consider an uncertainty $\epsilon$ in the full width at half maxima (FWHM),
and assume that the uncertainties at all iso-height contours of the beam
are a fixed fraction of the contour sizes, i.e.,
\begin{equation}
  \label{epsilon}
  \frac{d{\bf x}}{\bf x} = \epsilon.
\end{equation}
This uncertainty in the beam shape will then be transfered
to the multipole space
as the uncertainty in $\ell$ at a given height 
$B_{\ell m}$: 
\begin{equation}
  \label{dell}
  \frac{d\ell}{\ell} = -\epsilon.
\end{equation}
This results in the uncertainty in $B_{\ell}^2$ at a given $\ell$
\begin{equation}
  \label{Delta_Bl2}
  \Delta_{B_\ell^2} 
    =  \frac{dB_\ell^2}{B_\ell^2}
    =  \frac{B_{(1+\epsilon)\ell}^2}{B_\ell^2} -1.
\end{equation}
We then consider the change in the $C_\ell$ estimates:
\begin{equation}
  \label{Delta_Cl0}
  \Delta_{C_\ell}
    =  \frac{dC_\ell}{C_\ell}.
\end{equation}
Since the beam convolution occurs as the multiplication of
$B_\ell^2$ and $C_\ell$ (see eqs.~[\ref{ClBB}] and [\ref{ClBB_app}];
here we have dropped the subscript `(eff)' for concise notation),
we know that the resulting uncertainty in $C_\ell$ is
\begin{equation}
  \label{Delta_Cl}
  \Delta_{C_\ell}
    = \Delta_{B_\ell^2}
    = \frac{B_{(1+\epsilon)\ell}^2}{B_\ell^2} -1.
\end{equation}
This means that
if the beam size is mis-estimated by $\epsilon$
(i.e., the actual size is $1+\epsilon$ times the measured size),
then the resulting $C_\ell$ estimates will be $1+\Delta_{C_\ell}$ times
the real $C_\ell$.
Thus for a given uncertainty in the beam measurement $\epsilon$,
one can employ equation (\ref{Delta_Cl}) to estimate the resulting
uncertainty in the final $C_\ell$ estimates.
We also note that
the banding of $\ell$ does not affect this result,
as we shall show in section~\ref{uncertainties-from-beam-measurement-num}.

We note from the above result that
the type of uncertainty we assumed in the measurement
of the beam size (eq.~[\ref{epsilon}]) induces an error that is  
correlated between the $C_\ell$ estimates in all $\ell$ bins, 
although the magnitude of the error depends on $\ell$. 
If the power of the beam $B_\ell^2$ monotonically decreases with $\ell$,
a measured beam size slightly larger than the real value will produce
larger $C_\ell$ estimates at all $\ell$ bins. Although this is 
not the most general kind of beam size error, it is quite
common. 

We now investigate certain special cases.
In situations where
$\partial B_\ell^2 / \partial\ell$ is not changing much within $d\ell$, i.e.,
\begin{equation}
  \label{Delta_Bl2app_cond}
  \frac{\partial B_\ell^2}{\partial\ell}
  \approx 
  \frac{\partial B_{(1+\epsilon)\ell}^2}{\partial\ell},
\end{equation}
we can approximate equation (\ref{Delta_Bl2}) and thus (\ref{Delta_Cl}) as
\begin{equation}
  \label{Delta_Clapp}
  \Delta_{C_\ell} 
  =
  \Delta_{B_\ell^2} 
    \approx
      -\frac{d\ell}{B_\ell^2}\frac{\partial B_\ell^2}{\partial\ell}
    =  \frac{2\epsilon\ell}{B_\ell} \frac{\partial B_\ell}{\partial\ell},
\end{equation}
where equation (\ref{dell}) has been employed.
For a symmetric Gaussian beam $B({\bf x})=\exp(-x^2/2\varrho^2)$,
equation (\ref{Delta_Clapp}) becomes
\begin{equation}
  \label{Delta_Cl_Gau}
  \Delta_{C_\ell{\rm (G)}}
  \approx
  \Delta_{C_\ell{\rm (G)}}^*
  =
    - 2 \epsilon \varrho^2 \ell^2,
\end{equation}
while the condition (\ref{Delta_Bl2app_cond}),
for $|\epsilon|\ll 1$, leads straightforwardly to
\begin{equation}
  \label{Delta_Bl2app_cond2}
  \ell \ll \ell_{\rm(G)}^* = \frac{1}{\sqrt{2|\epsilon|}\varrho}.
\end{equation}
Here we have again used the small-field limit.
When combined with the condition (\ref{Delta_Bl2app_cond2}),
we find that 
approximation (\ref{Delta_Cl_Gau}) breaks down
when $|\Delta_{C_\ell{\rm (G)}}^*|$ is comparable with unity.
In particular,
we investigate the accuracy of approximation 
(\ref{Delta_Cl_Gau}),
by comparing it with equation (\ref{Delta_Cl}).
We find for $|\epsilon|<20\%$ that 
the approximation is accurate within $10\%$ error
if
\begin{equation}
  \label{lcond}
  \ell < \ell_{{\rm(G:10\%)}}^* = (0.44+0.8|\epsilon|)\; \ell_{{\rm(G)}}^*,
\end{equation}
where $\ell_{{\rm(G)}}^*$ is given in equation (\ref{Delta_Bl2app_cond2}).
For example,
if $\epsilon=10\%$ and 
the Gaussian beam has a FWHM of 10 arcminutes 
(i.e., $\varrho=1.24\times 10^{-3}$ radians),
then approximation (\ref{Delta_Cl_Gau}) is accurate
within $10\%$ error when $\ell < \ell_{{\rm(G:10\%)}}^* \approx 940$.
Under the condition (\ref{lcond}),
one can see from equation (\ref{Delta_Cl_Gau}) that,
for an approximately Gaussian beam,
the resulting uncertainty in the final $C_\ell$ estimates
increases in proportion to the uncertainty in the beam measurement $\epsilon$,
the square of the beam size $\varrho^2$,
and the square of the multipole number $\ell^2$.

\section{DECONVOLUTION OF THE PIXEL SMOOTHING}
\label{deconvolution_of_the_pixel_smoothing}

We have not dealt with
the smoothing effects due to the pixelization of the map,
when translating the data from the temporal to the pixel domain
(see eqs.~[\ref{m}] and [\ref{BplmY0}]).
Because 
convolving a CMB map with a Dirac Delta
${\delta}({\bf x}-{\bf x}_1)$
will shift the original temperature at ${\bf x}$
to a new location ${\bf x}+{\bf x}_1$,
we know that 
$Y_{\ell m}({\bf x})=
{\delta}_{\ell m}({\bf x}_1) Y_{\ell m}({\bf x}+{\bf x}_1)$
where ${\delta}_{\ell m}({\bf x}_1)$ 
is the multipole expansion of 
${\delta}({\bf x}-{\bf x}_1)$.
This allows us to rewrite equation (\ref{BplmY0}) as
\begin{equation}
  \label{Bplm0}
  B_{p\ell m} = C_N A^T N^{-1} B_{t \ell m}
  {\delta}_{\ell m}({\bf x}_p-{\bf x}_t).
\end{equation}
Substituting this into equation (\ref{barBplm00}),
we obtain
\begin{equation}
  \label{barBplm0}
  \overline{B}_{p\ell m} = 
  H U^T C_{N}^{-1} {\rm diag}(C_{N} A^T N^{-1} B_{\delta\ell m}),
\end{equation}
where 
$B_{\delta\ell m}\equiv B_{t \ell m}{\delta}_{\ell m}({\bf x}_p-{\bf x}_t)$
is a ${\cal N}_t$ by ${\cal N}_p$ matrix,
and ${\rm diag}(M)$ is a vector
whose entries are the diagonal elements of the matrix $M$.
These results are completely general.
Without further information about $N^{-1}$ or $B_{\delta\ell m}$,
equation (\ref{barBplm0}) can not be simplified, 
mainly due to the involvement of ${\delta}_{\ell m}({\bf x}_p-{\bf x}_t)$.

With the white-noise assumption (see sec.\ \ref{the_pixel_beam_expansion}),
we have equation (\ref{Bplm0}) simplified as
\begin{equation}
  \label{Bplmpi}
  B_{\Pi p\ell m}
  = \sum_{t\in p}\xi_t B_{t\ell m}{\delta}_{\ell m}({\bf x}_p-{\bf x}_t),
\end{equation}
and equation (\ref{barBplm0}) as
\begin{equation}
  \label{barBplm2pi}
  \overline{B}_{\Pi p\ell m}
  = \sum_t \chi_t B_{t\ell m}{\delta}_{\ell m}({\bf x}_{p\ni t}-{\bf x}_t),
\end{equation}
where 
${\bf x}_{p\ni t}$ is the central coordinates of the pixel $p$
that covers ${\bf x}_t$,
and
$\xi_t$ and $\chi_t$ are as defined in equations
(\ref{xi_t}) and (\ref{chi_t}) respectively.
Here we use the subscript `$\Pi$' to distinguish these results
from those in equations (\ref{Bplm}) and (\ref{barBplm2}).
In the real space,
equation (\ref{barBplm2pi}) is equivalent to 
\begin{equation}
  \label{barBp2pi}
  \overline{B}_{\Pi p} ({\bf x})
  = \sum_t \chi_t B_{t}({\bf x}-{\bf x}_{p\ni t}+{\bf x}_t),
\end{equation}
meaning that
$\overline{B}_{\Pi p}$
is the noise-weighted average 
over the time-stream beams $B_{t}$ 
that are shifted by ${\bf x}_{p\ni t}-{\bf x}_t$ at each time $t$.
This implies that
our formalism developed previously is still available,
requiring only a modification
that takes into account
the detailed locations of the temporal hits ${\bf x}_t$
with respect to the pixel centers ${\bf x}_{p\ni t}$.
Thus we have relaxed the assumption ${\bf x}_t\equiv {\bf x}_{p\ni t}$
that was made to achieve equations (\ref{Bplm}) and (\ref{barBplm2}).

In most cases,
both ${\cal N}_p$ and ${\cal N}_t$ are large,
and the beam shape $B_t$ of each photometer does not change much 
within several successive pixels.
This results in the fact that
in determining the $\overline{B}_{\Pi p}$ 
in equation (\ref{barBp2pi}),
each beam configuration $A(\beta)B_0({\bf x})$ 
(see eq.~[\ref{B0}])
appears at a set of ${\bf x}_t$
which have offsets ${\bf x}_{p\ni t}-{\bf x}_t$
distributed within a region
that is confined by the pixel shapes.
If all pixels have the same shape, 
then
this is equivalent to
convolving each $A(\beta)B_0({\bf x})$ with a top-hat like window
whose boundary is defined by the pixel shape.
As a result,
we can approximate equation (\ref{barBplm2pi}) as
\begin{equation}
  \label{barBplm2piapp}
  \overline{B}_{\Pi p\ell m}
  \approx \overline{B}_{p\ell m} \Pi_{\ell m},
\end{equation}
where $\overline{B}_{p\ell m}$ is as defined in equation (\ref{barBplm2}),
and $\Pi_{\ell m}$ is the multipole expansion of
\begin{equation}
  \label{Pix}
  \Pi({\bf x})=\sum_{t} {\delta} ({\bf x}_{p\ni t}-{\bf x}_t).
\end{equation}
The same also applies to the simple case
where all time-stream beams $B_t$ are the same.
We thus see that
the $\Pi_{\ell m}$ in equation (\ref{barBplm2piapp})
serves as an extra convolution 
(apart from the time-stream beam convolution)
of the CMB signal due to the pixelization of the map.
With such,
we can now easily incorporate this extra smoothing effects into our formalism
by replacing our $B_{\ell{\rm (eff)}}^2$ with
(see also eq.~[\ref{Bl2eff_avg4}])
\begin{eqnarray}
  B_{\Pi \ell{\rm (eff)}}^2
  &  \approx &
  \overline{B}_{\Pi p\ell{\rm (ms)}}^2
  \label{Bl2_effx} \\
   &\approx  &
  \overline{B}_{p\ell{\rm (ms)}}^2
  \Pi_{\ell{\rm (ms)}}^2
  \label{Bl2_effxapp} \\
   &\approx  &
  B_{\ell{\rm (eff)}}^2
  \Pi_{\ell{\rm (ms)}}^2,
  \label{Bl2_effxapp2} 
\end{eqnarray}
where 
$\overline{B}_{\Pi p\ell({\rm ms})}^2$ is the power spectrum of
the $\overline{B}_{\Pi p}$ defined in equation (\ref{barBp2pi}),
and $\Pi_{\ell{\rm (ms)}}^2$ is the power spectrum of the $\Pi({\bf x})$
defined in equation (\ref{Pix}). 
We note that
in the limiting case where ${\bf x}_t={\bf x}_{p\ni t}$,
we have $\Pi_{p\ell m}$ equal to unity for all $\ell$ and $m$
(since the multipole transform of $\delta({\bf x})$ is unity),
so the smoothing effects disappear,
and we have exactly $B_{\Pi \ell{\rm (eff)}}^2=B_{\ell{\rm (eff)}}^2$
(see eq.~[\ref{Bl2_effxapp2}]).
If the pixels do not have exactly the same shape, as in the case on
any large patch of the sphere (e.g., pixelized by
HEALPix, Gorski et al.~1999,
or
by Igloo, Crittenden \& Turok 1998),
then we can use equation (\ref{Bl2_effx}) together with equation (\ref{barBp2pi}) 
to obtain $B_{\Pi \ell{\rm (eff)}}^2$.
If the pixel beam or the pixel shape remains roughly the same for all pixels,
then we can use equation (\ref{Bl2_effxapp}) 
together with equations (\ref{Bl2eff_avg4}) and (\ref{Pix})
to calculate $B_{\Pi \ell{\rm (eff)}}^2$.

If all the pixels have the same shape
which is a regular square of size $\varsigma$ in radians,
then we have
\begin{equation}
  \label{Bl2_pxl}
  \Pi_{\ell{\rm (ms)}}^2(\varsigma)
  = \frac{8}{\pi}\int_0^{2\pi}d\phi
    \frac{\sin^2\left[\left({\ell\varsigma\cos\phi}\right)/{2}\right]
          \sin^2\left[\left({\ell\varsigma\sin\phi}\right)/{2}\right]}
         {\ell^4\varsigma^4(\cos^2\phi)(\sin^2\phi)}.
\end{equation}
An accurate approximation to this result is
\begin{equation}
  \label{Bl2_pxlapp}
  \Pi_{\ell{\rm (ms)}}^2(\varsigma)
  \approx
  \exp\left[-\frac{(\ell\varsigma)^{2.04}}{18.1}\right]
  \left[1-2.72\times 10^{-2}(\ell\varsigma)^2\right].
\end{equation}
The accuracy of this fit is within $0.3\%$ error for
$\ell<1.4 \pi/\varsigma$.
For example,
if the pixel size is $5\times 5$ square arcminutes 
(i.e., $\varsigma=$ 5 arcminutes $\approx 1.45\times 10^{-3}$ radians),
then the above fit is at $99.7\%$ accuracy
for $\ell<3024$.

\section{NUMERICAL VERIFICATIONS}
\label{numerical_verifications}

\subsection{The pixel-beam expansion}
\begin{figure}
  \epsscale{0.5}
  \plotone{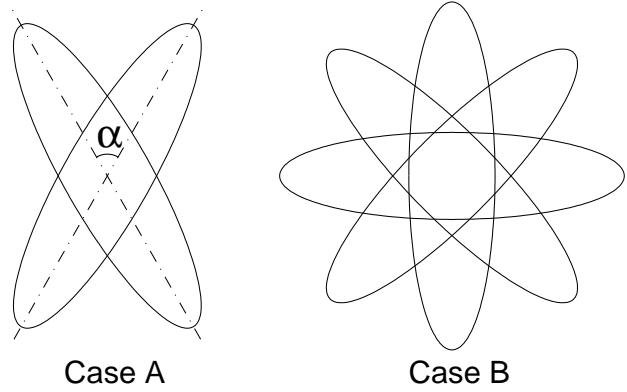}
  \caption
  {Beam configurations on a given pixel,
resulting from different scanning strategies.}
  \label{fig:beam_config}
\end{figure}

In this and the following three subsections,
we will employ an elliptic Gaussian beam
with a short-axis FWHM of 5 arc-minutes
and a long-axis FWHM of 20 arc-minutes,
to demonstrate certain key points developed previously.
We first investigate the pixel-beam expansion of a given pixel
resulting from different scanning strategies, i.e., 
to investigate the dependence of the pixel-beam power spectrum
(\ref{Bl2eff_3}) on the function $f(\beta)$,
and to verify the results given in equation (\ref{B2l_constraint}).
We note that
although
those results are given for the average pixel-beam expansion,
we expect the pixel-beam expansion to carry the same property
since equation (\ref{Bplm}) has exactly the same form as 
equation (\ref{barBplm2}).
Figure~\ref{fig:beam_config}
 shows two different configurations
of beam scanning on a given pixel.
In case A, the pixel was hit twice by the same beam pattern,
but with different orientations of a separation angle $\alpha$.
That is $f(\beta)\equiv[\delta(\beta)+\delta(\beta-\alpha)]/2$.
We shall investigate the cases $\alpha=15$, $45$, and $90$ degrees.
In case B, the pixel was hit evenly in four different directions.
That is $f(\beta)\equiv
[\delta(\beta)+\delta(\beta-45^\circ)
+\delta(\beta-90^\circ)+\delta(\beta-135^\circ)]/4$.

\begin{figure}
  \epsscale{0.5}
  \plotone{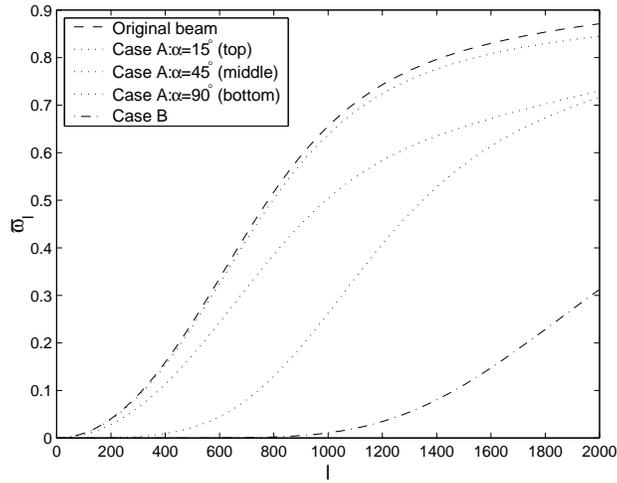}
  \caption
  {Indices of asymmetry of the pixel-beam expansions,
as functions of the multipole number $\ell$.}
  \label{fig:wl_pxl}
\end{figure}

Figure~\ref{fig:wl_pxl}
 shows the IOA of the pixel beams in cases A and B,
as defined in equation (\ref{varpi_l}).
As one can see,
the pixel beam has the largest asymmetry (largest $\varpi_\ell$)
when the pixel is hit by a beam with only one direction (the dashed line).
When the pixel is hit by beams of two different directions
(case A in Figure~\ref{fig:beam_config}
),
the asymmetry decreases
($\varpi_\ell$ decreases)
if the separation angle of the two directions $\alpha$ is closer to 90 degrees
(see the dotted lines in Figure\ref{fig:wl_pxl}
).
When the pixel is scanned with four different directions 
(case B in Figure~\ref{fig:beam_config}
),
the resulting effective beam is nearly symmetric ($\varpi_\ell\approx 0$)
up to $\ell\sim 1000$,
and has the lowest level of asymmetry (the smallest $\varpi_\ell$).

\begin{figure}
  \epsscale{0.5}
  \plotone{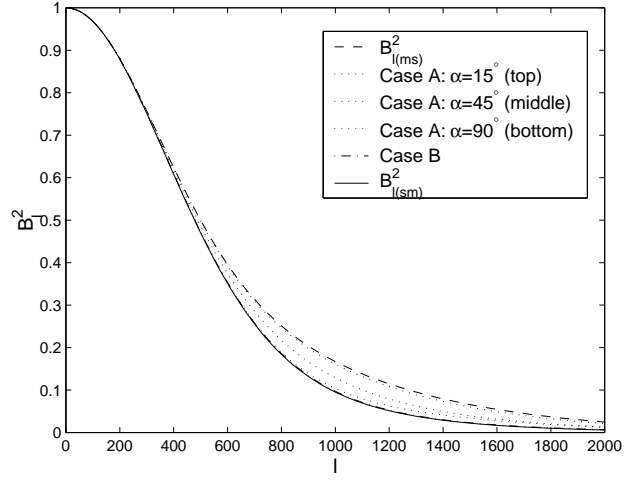}
  \caption
  {Power spectra $B_\ell^2$ of the pixel-beam expansions
  with different scanning strategies.}
  \label{fig:Bl2_pxl}
\end{figure}

Figure~\ref{fig:Bl2_pxl}
 shows the power spectra of the pixel-beam expansions
with different scanning strategies.
As we can see,
the power spectrum of the pixel-beam expansion has a maximum
given by equation (\ref{B2lms}) (see also eq.~[\ref{Bl2eff_case2}]),
when the pixel was scanned with only one direction.
On the other hand,
the power spectrum of the pixel-beam expansion has a minimum
given by equation (\ref{B2lsm}) (see also eq.~[\ref{Bl2eff_case1}]),
when the pixel was scanned evenly in all directions
(note that the dot-dashed line in Figure~\ref{fig:Bl2_pxl}
almost coincides with the solid line).
This verifies our results given in equation (\ref{B2l_constraint}).
By comparing Figure~\ref{fig:Bl2_pxl}
 with Figure~\ref{fig:wl_pxl},
we also learn that
there is a strong correlation between $\varpi_\ell$ and the
$B_\ell^2$ of a pixel---when the pixel 
is scanned by a same beam pattern with more different directions,
the level of the effective beam asymmetry ($\varpi_\ell$) decreases,
and so does the power spectrum of the pixel-beam expansion ($B_\ell^2$).

We also note that
according to equation (\ref{Clmid_constraint}),
the IOA of the original time-stream beam $B_0$ (the dashed line
in Fig.~2) also tells us beyond what $\ell$
we need to worry about the asymmetry of the beam.
For example,
when $\ell\lesssim 600$,
we see that $\varpi_\ell\lesssim 0.3$,
giving $\varpi_\ell^2/2\lesssim 0.05$.
This means that
if we simply use the 
$B_{0\ell{\rm (mid)}}^2=2/[B_{0\ell{\rm (ms)}}^{-2}+B_{0\ell{\rm (sm)}}^{-2}]$ 
(eq.~[\ref{B20lmid}]) 
as the $B_{\ell{\rm (eff)}}^2$
in the formalism (\ref{ClBB_app}),
then the maxima error in the final $C_\ell$ estimates
is guaranteed to be within about $\pm 5\%$ for $\ell\lesssim 600$.

\subsection{The pixel-pixel beam expansion}
\label{the-effective-two-point-beam-expansion-num}

\begin{figure}
  \epsscale{0.5}
  \plotone{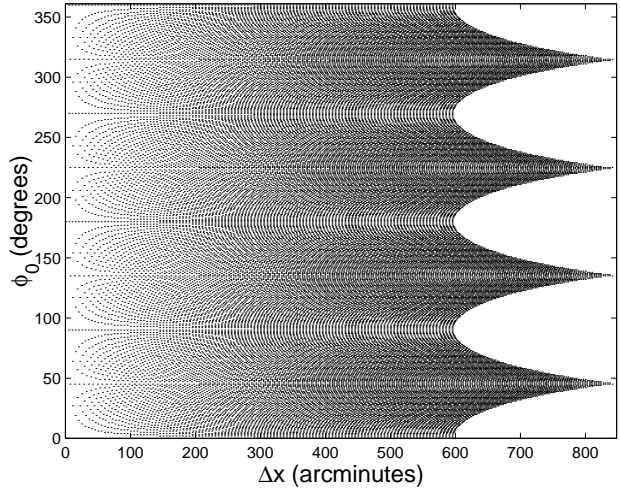}
  \caption
  {Distribution of $\varphi_0$ as a function of $\Delta x$.
  Each dot labels the $(\Delta x,\varphi_0)$
  that is sampled by a square map of $10^\circ\times 10^\circ$,
  with a pixel size of 5 arcminutes.}
  \label{fig:dx_phi0}
\end{figure}

We now use the elliptic Gaussian beam of 5 by 20 arcminutes
to verify some important results 
in section~\ref{the_effective_two_point_Bl}---mainly
equation (\ref{Bl2eff_avg4}).
Consider a square map of size $10^\circ\times 10^\circ$,
with a square pixel size of 5 arcminutes.
Referring to equation (\ref{Bl2eff_avg2})
with such a map,
Figure~\ref{fig:dx_phi0} shows
how $\varphi_0$ is distributed at each $\Delta x$.
In the figure, each dot labels the $(\Delta x,\varphi_0)$
that is sampled by the map.
As one can see,
$\varphi_0$ is nearly uniformly distributed for any given $\Delta x$,
except when $\Delta x$ is close to the boundaries constrained by the
pixel and field sizes.
Because of this nearly uniform distribution,
we achieved equation (\ref{Bl2eff_avg4}) from equation (\ref{Bl2eff_avg2}).

More precisely,
we carried out equation (\ref{Bl2eff_avg2}) to obtain $B_{\ell{\rm(eff)}}^2$,
and calculated the right hand side of equation (\ref{Bl2eff_avg4})
to obtain $B_{\ell{\rm(ms)}}^2$.
Here we have used the elliptic Gaussian beam directly as
the $\overline{\widetilde{B}}_p({\bf k})\equiv \overline{B}_{p\ell m}$
in the due calculations.
We found that
$B_{\ell{\rm(ms)}}^2$ agrees with $B_{\ell{\rm(eff)}}^2$
with more than $99\%$ accuracy for $\ell=0$--$2000$.

\begin{figure}
  \epsscale{0.5}
  \plotone{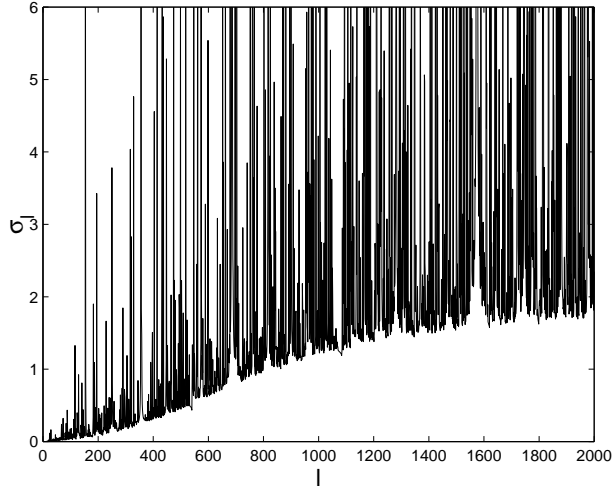}
  \caption
  {Mean discrepancy $\sigma_\ell$
  of $B_{pp'\ell}^2$ from $B_{\ell{\rm(eff)}}^2$
  for each individual pixel pair.}
  \label{fig:sig_l}
\end{figure}

We have also calculated the average deviation $\sigma_\ell$
of $B_{pp'\ell}^2$ from $B_{\ell{\rm(eff)}}^2$
for each individual pixel pair,
using equation (\ref{sigmal2}).
The result is shown in Figure~\ref{fig:sig_l}.
First,
we see many spikes in $\sigma_\ell$.
This is due to the zeros of the Bessel function $J_0$,
which appears at the bottom of equation (\ref{sigmal2}).
These spikes should be neglected, as in reality
no such singularities appear in our analysis pipeline.
We note that
these spikes have the same origin as those presented in
Hanany et~al.~(1998),
where a similar situation was considered.
Second,
as addressed previously,
although the $\sigma_\ell$ obtained from equation (\ref{sigmal2}) can
be as large as comparable to unity,
the real errors in the final $C_\ell$ estimates
by using the formalism (\ref{CBBB_app3})
with the approximation (\ref{Bl2eff_avg4})
will be much smaller than this value.
This is because
the $\sigma_\ell$ here tells only 
the mean discrepancy of $B_{\ell{\rm(eff)}}^2$ for each individual pixel pair,
and
may average out when all pixel pairs come into account
in the likelihood analysis.
In section~\ref{the-maxima},
we will numerically justify this and thus
the accuracy of employing equation (\ref{CBBB_app3}) with (\ref{Bl2eff_avg4}).

\subsection{Uncertainties from beam measurement}
\label{uncertainties-from-beam-measurement-num}

\begin{figure}
  \epsscale{0.5}
  \plotone{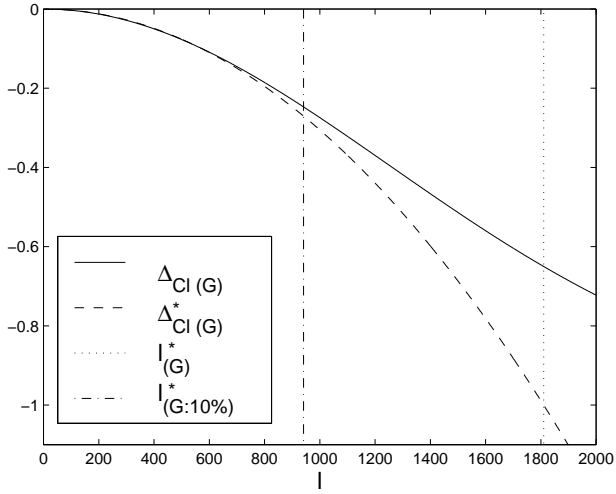}
  \caption
  {Uncertainty in the $C_\ell$ (solid line; given by eq.~[\ref{Delta_Cl}]),
  resulting from an uncertainty of $\epsilon=10\%$ in the beam shape measurement
  of a Gaussian beam with a FWHM of 10 arcminutes.
  Also plotted is the approximation (\ref{Delta_Cl_Gau}) (dashed line).
  The dotted vertical line indicates the limit of the approximation
  $\ell_{{\rm(G)}}^*$ given by equation (\ref{Delta_Bl2app_cond2}),
  while the dot-dashed vertical line shows
  the 10\% accuracy limit 
  $\ell_{{\rm(G:10\%)}}^*$,
  which is well fitted by equation (\ref{lcond}).}
  \label{fig:Delta_sym}
\end{figure}

In this section,
we will numerically verify the results in 
section~\ref{uncertainties_from_beam_measurement}.
First,
we use a symmetric Gaussian beam with a FWHM of 10 arcminutes,
to investigate the approximation $\Delta_{C_\ell {\rm (G)}}^*$
given by (\ref{Delta_Cl_Gau}),
as a comparison to the exact result $\Delta_{C_\ell {\rm (G)}}$
given by (\ref{Delta_Cl}).
Here we take the uncertainty in the beam measurement
to be $\epsilon=10\%$ (eq.~[\ref{epsilon}]).
As one can see in Figure~\ref{fig:Delta_sym}, 
the approximation breaks down towards the limit 
$\ell_{{\rm(G)}}^*$ given by equation (\ref{Delta_Bl2app_cond2}).
For $\ell\ll \ell_{{\rm(G)}}^*$,
the approximation $\Delta_{C_\ell {\rm (G)}}^*$ reproduces
the correct result $\Delta_{C_\ell {\rm (G)}}$.
By comparing $\Delta_{C_\ell {\rm (G)}}^*$ and $\Delta_{C_\ell {\rm (G)}}$,
we calculate the $10\%$ accuracy limit $\ell_{{\rm(G:10\%)}}^*$
(the dot-dashed line),
at which $\Delta_{C_\ell {\rm (G)}}^*/\Delta_{C_\ell {\rm (G)}}-1=10\%$.
In addition,
by varying the value of $\epsilon$ between $\pm 20\%$,
we obtain the result presented in equation (\ref{lcond}).
That is,
for a symmetric Gaussian beam with an uncertainty of $\epsilon$ in size,
the approximation (\ref{Delta_Cl_Gau}) for the resulting uncertainty in $C_\ell$
is accurate within $10\%$ error
for $\ell< \ell_{{\rm(G:10\%)}}^*=(0.44+0.8|\epsilon|) \ell_{{\rm(G)}}^*$.

\begin{figure}
  \epsscale{0.5}
  \plotone{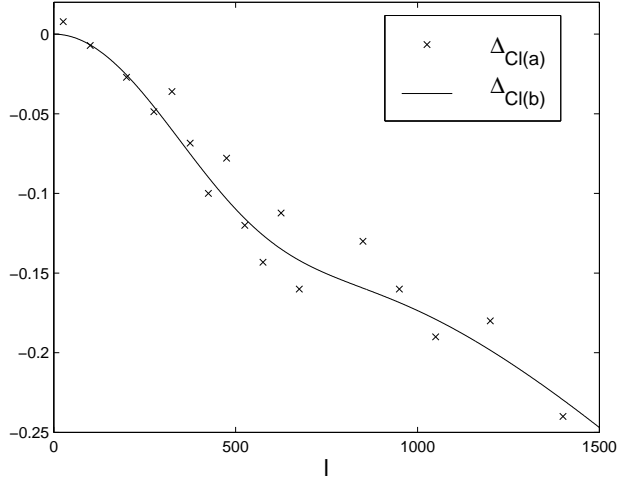}
  \caption
  {Uncertainty in the $C_\ell$ resulting from that in the beam measurement.
  The horizontal axis is the multipole number $\ell$.
  The crosses, $\Delta_{C_\ell{\rm(a)}}$, are results based on simulations
  using equation (\ref{Delta_Cl0}) (or eq.~[\ref{Delta_Cl_sim}]),
  while the solid line, $\Delta_{C_\ell{\rm(b)}}$,
  is obtained directly from the beam shape
  using equation (\ref{Delta_Cl}).
  An elliptic Gaussian beam,
  with long- and short-axis FWHM's of 20 and 5 arcminutes respectively,
  is used.
  The uncertainty in the beam shape is $\epsilon=10\%$.}
  \label{fig:D_Cl}
\end{figure}

Now we investigate the case where the beam is asymmetric.
We use an elliptic Gaussian beam,
whose long- and short-axis FWHM's are 20 and 5 arcminutes respectively.
This beam is first convolved onto a simulated CMB map of size 
$10^\circ \times 10^\circ$,
with a pixel size of 10 arcminutes.
The underlying cosmology is an inflationary model with
$(\Omega_{\rm b},\Omega_{\rm cdm},\Omega_\Lambda,n,h)=(0.07,0.61,0.23,1,0.60)$,
normalized to the COBE DMR.
A random Gaussian noise of $100\mu K$ is then added into each pixel.
We call this simulation (1).
We repeat the same procedure again except that this time 
the beam size is increased by $10\%$, i.e., $\epsilon=10\%$,
to obtain a simulation (2),
where the CMB and noise realizations are exactly the same as those used
in simulation (1).
We then analyze both simulations
using the procedure outlined in section~\ref{the_convention},
with the approximation (\ref{CBBB_app3}).
The resulting uncertainty in $C_\ell$ can thus be calculated using
equation (\ref{Delta_Cl0}) as
\begin{equation}
  \label{Delta_Cl_sim}
  \Delta_{C_\ell{\rm(a)}}
    =  \frac{C_{\ell(2)}}{C_{\ell(1)}}-1,
\end{equation}
where the subscripts (1) and (2) indicate results from the two simulations.
The results are shown as crosses in Figure~\ref{fig:D_Cl}.
Also plotted is the result using equation (\ref{Delta_Cl}) (the solid line),
which we label with a subscript (b).
It is obtained directly by varying the beam shape with $\epsilon=10\%$.
As one can see,
the crosses are highly consistent with the solid line.
This means, first, that
the asymmetry of the beam does not affect 
our result given by equation (\ref{Delta_Cl}).
Second,
the banding of $\ell$ does not affect the result,
so we can use equation (\ref{Delta_Cl})
as an estimate for the uncertainty in the band power $C_b$
resulting from that in the beam measurement.
This is also an important support to the fact that
the banding of $\ell$ does not affect 
the general relation $C_\ell\propto B_\ell^{-2}$
(see eqs.~[\ref{ClBB_app}] and [\ref{CBBB_app3}]).
We have also verified that
the sizes of the error bars in the $C_\ell$ estimates 
between simulations (1) and (2)
do not change by more than $4\%$ for $\ell<1200$.
Thus we know that 
when $\Delta_{C_\ell}$ is small,
the uncertainty in the beam shape measurement does not affect
the sizes of error bars significantly,
but does affect the amplitudes of the $C_\ell$ estimates.
On the other hand,
when $\Delta_{C_\ell}$ is large (comparable to one),
the signal to noise ratio may be affected and so may the error bar sizes.

\subsection{Deconvolution of the pixel smoothing}
\label{deconvolution_of_the_pixel_smoothing-num}

\begin{figure}
  \epsscale{0.5}
  \plotone{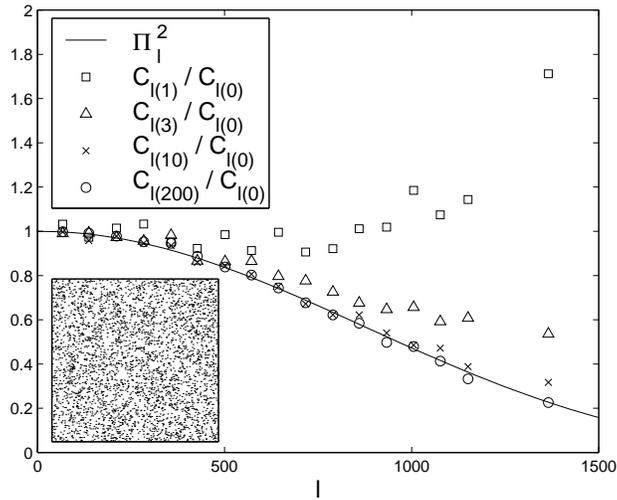}
  \caption
  {Smoothing effect due to the pixelization of a CMB map.
  The ratios $C_{\ell(j)}/C_{\ell(0)}$ are compared to the approximated
  smoothing window $\Pi_{\ell{\rm (ms)}}^2$, 
  with $j=1$, 3, 10, 200 representing the number of temporal samples per pixel
  in different runs.
  The horizontal axis is the multipole number $\ell$.
  The square at the bottom-left corner shows the $\Pi({\bf x})$ for $j=10$.}
  \label{fig:pxl_smth}
\end{figure}

We now test the formalism
of deconvolving the smoothing effect due to the pixelization of a map.
This is to verify equation (\ref{Bl2_effxapp}),
with equation (\ref{Bl2_pxlapp}) as an approximation in cases
where the pixels are regular squares.
We consider a square CMB map of size $10^\circ \times 10^\circ$,
with regular-square pixels of size 10 arcminutes.
We first simulate a time-stream of the CMB signal $\gamma_t$,
that is convolved with an elliptic Gaussian beam 
of 5 by 20 arcminutes in FWHM 
(same as the one used in previous sections).
For each temporal sample,
we then add Gaussian random white noise $n_t$ with $5\%$ in RMS amplitude.
In this run,
we require the temporal samples to be exactly at the centers of each pixels,
i.e., ${\cal N}_t = {\cal N}_p$ and ${\bf x}_t={\bf x}_{p\ni t}$,
such that $m=d$ (see eq.~[\ref{m}]).
We call this simulation (0).
In a second run,
the procedure is the same except that the CMB temporal samples now have
offsets with respect to the centers of each pixels, 
i.e.,  ${\cal N}_t = {\cal N}_p$ 
but ${\bf x}_t \neq {\bf x}_{p\ni t}$ with  ${\bf x}_t-{\bf x}_{p\ni t}$
randomly distributed within a square of size 10 arcminutes.
We call this simulation (1).
In third, fourth, and fifth runs,
the procedures are the same as simulation (1),
except that
the numbers of temporal samples in each pixels are now
3, 10, and 200 
(i.e., ${\cal N}_t = 3{\cal N}_p$, $10{\cal N}_p$, and  $200{\cal N}_p$)
respectively, instead of one.
We denote these as simulations (3), (10), and (200) respectively.
All these runs are then analyzed in the same way,
using the procedure outlined in section~\ref{the_convention},
with the approximation (\ref{CBBB_app3})
and $B_{\ell{\rm (eff)}}^2=\overline{B}_{p\ell{\rm (ms)}}^2$
(eq.~[\ref{Bl2eff_avg4}]).
Therefore the ratio
\begin{equation}
  \label{Ci0}
  \frac{C_{\ell(j)}}{C_{\ell(0)}}, \quad j=1,\; 3,\; 10,\; 200,
\end{equation}
will quantify the smoothing effect due to the pixelization of the map.
We plot this ratio in Figure~\ref{fig:pxl_smth},
as a comparison to the $\Pi_{\ell{\rm (ms)}}^2$ given in equation (\ref{Bl2_pxlapp}).

As one can see and expect,
the smoothing effect approaches the top-hat-window approximation
when the number of temporal samples per pixel increases.
When it is larger than 10, as in most real situations,
the top-hat-window approximation appears to be a good one.
Also plotted at the bottom-left corner is the $\Pi({\bf x})$
given by equation (\ref{Pix}) for $j=10$.
The nearly uniform distribution of ${\bf x}_{p\ni t}-{\bf x}_t$ shows
the appropriateness of the top-hat-window approximation.
Thus we have verified that
the approximation (\ref{Bl2_effxapp}),
with equation (\ref{Bl2_pxlapp}) for cases where pixels are regular squares,
is indeed a good approximation.
For an obvious mathematical reason
(see sec.~\ref{deconvolution_of_the_pixel_smoothing}),
we know that
this formalism can be further extrapolated 
for cases where pixels do not have regular shapes
but the time-stream beams have roughly the same shape.
In such cases,
equation (\ref{Pix}) can be employed to obtain the $\Pi_{\ell{\rm (ms)}}^2$
for the use of equation (\ref{Bl2_effxapp}).

\subsection{The MAXIMA experiment}
\label{the-maxima}

\begin{figure}
  \epsscale{0.5}
  \plotone{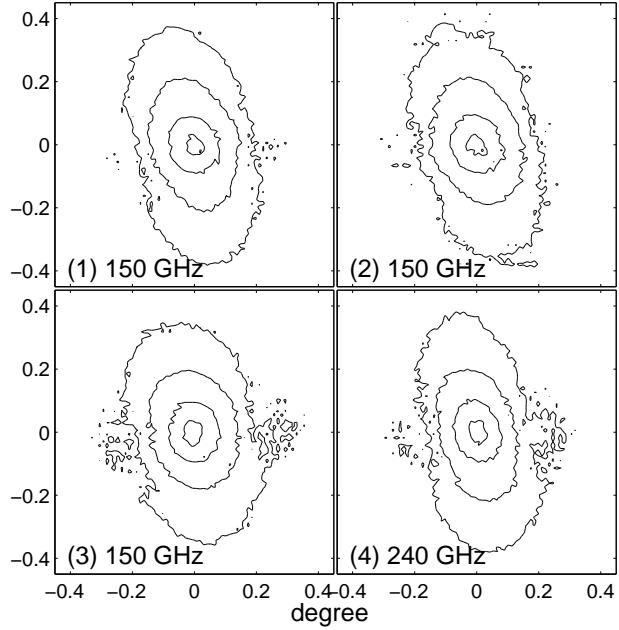}
  \caption
  {Iso-height contours of the beams used in the analysis 
  of the MAXIMA-1 data (H00).
  Contours correspond to the 
  90\%, 50\%, 10\%, and 1\% amplitude levels.}
  \label{fig:beamshapes}
\end{figure}

In this section we demonstrate the application of our formalism using
the data from the MAXIMA-1 experiment (H00).  
Figure~\ref{fig:beamshapes}
shows the antenna patterns $B_i$ ($i=1,2,3,4$)
 for the four photometers used in the
analysis of the MAXIMA-1 data. 
Details of the measurements of these
beam shapes are given in H00.
As one can see,
the beams are more
symmetric towards their centers.

\begin{figure}
  \epsscale{0.5}
  \plotone{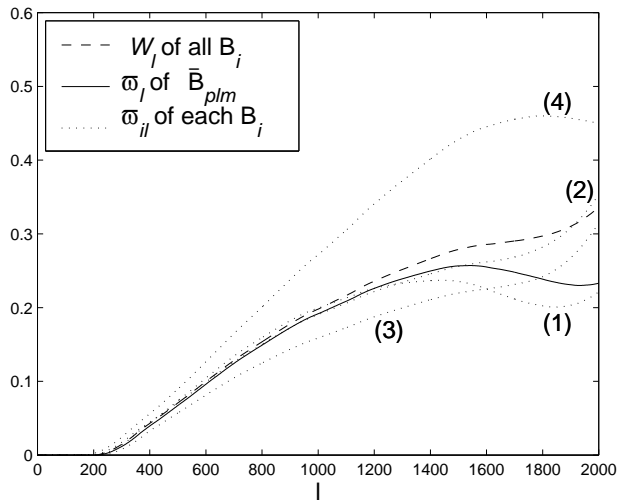}
  \caption
  {Indices of asymmetry of MAXIMA-1 beams (dotted lines) and 
  of the noise-weighted combination (solid line).
  Also plotted is the index of combined asymmetry of all MAXIMA-1 beams
  (dashed line).}
  \label{fig:wl_ma}
\end{figure}

Figure~\ref{fig:wl_ma} shows the level of asymmetry of the beams.
The dotted lines are the IOA $\varpi_{i\ell}$ of each individual beam $B_i$ 
(eq.~[\ref{varpi_l}]),
and the solid
line is the IOA of the noise-weighted combination of all of them,
i.e., 
the $\varpi_{\ell}$ of the average pixel-beam expansion $\overline{B}_{p\ell m}$
(see eqs.~[\ref{barBplm2}], [\ref{chi_t}], and [\ref{mu_t}]). 
The dashed line is the IOCA ${\cal W}_\ell$ from all $B_i$
(eq.~[\ref{W_l}]).
Here the relative weight of each beam is 
$\zeta_1:\zeta_2:\zeta_3:\zeta_4=64:64:81:36$ 
(see eq.~[\ref{zeta}] and H00).
As we can see,
the beams are nearly symmetric (${\cal W}_\ell, \varpi_\ell\approx 0$) at low
$\ell$, but less so at larger $\ell$.
At $\ell \lesssim 800$,
which is the range of $\ell$ discussed in H00,
the asymmetry is less than $15\%$.
The figure also confirms the fact that 
the $\varpi_{\ell}$ of $\overline{B}_{p\ell m}$ must be equal to or smaller than
${\cal W}_\ell$, although the individual $\varpi_{i\ell}$ may be larger than
${\cal W}_\ell$ (see sec.~\ref{the_criteria_for_beam_symmetry}).

\begin{figure}
  \epsscale{0.5}
  \plotone{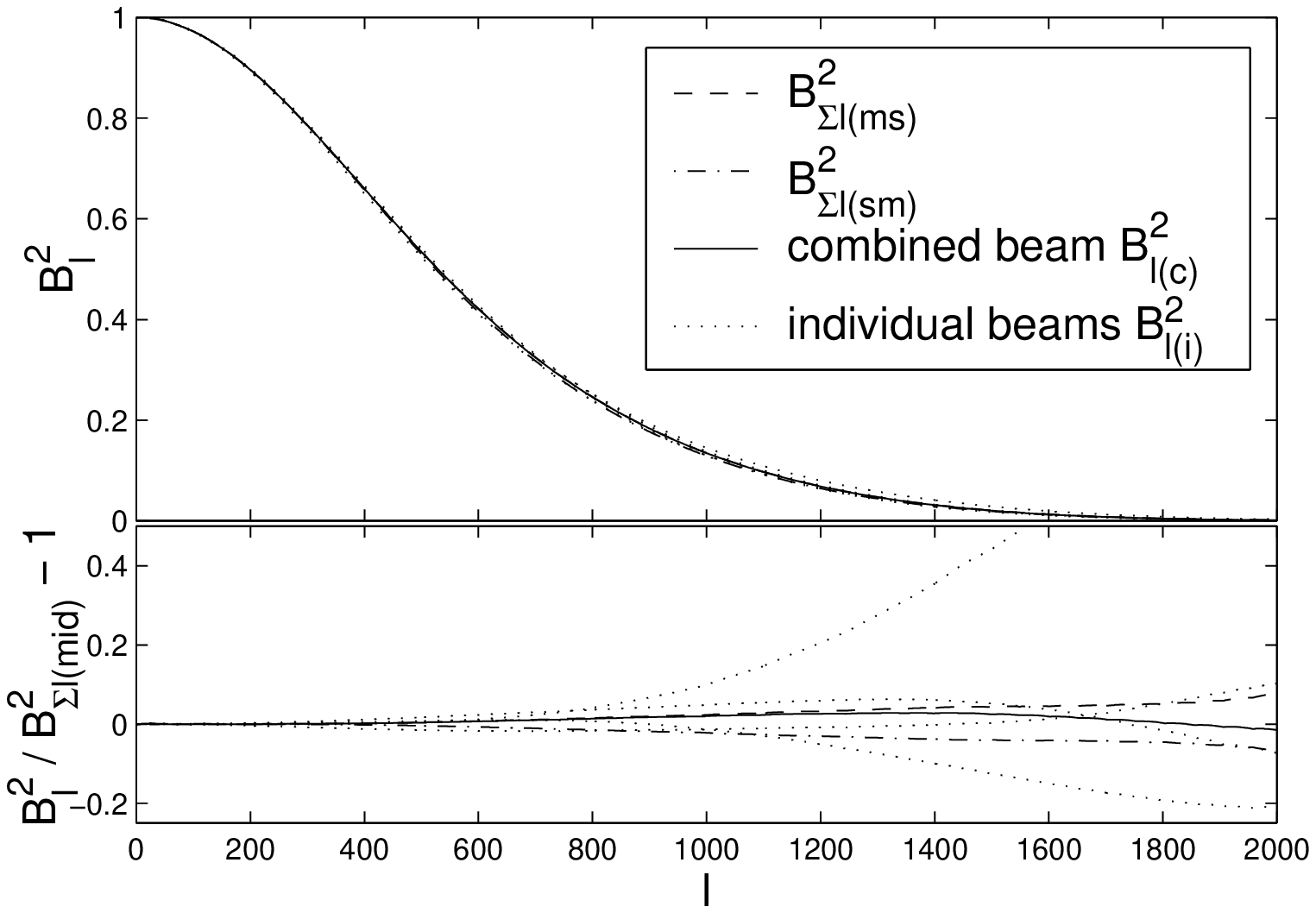}
  \caption
  {Pixel-pixel beam expansions $B_{\ell{\rm (eff)}}^2$
  of MAXIMA-1 beams and
  their noise weighted combination (top panel), and  
  a comparison of these results (bottom panel; see text). 
  Also plotted are
  the $B_{\Sigma\ell{\rm (sm)}}^2$
  and $B_{\Sigma\ell{\rm (ms)}}^2$.}
  \label{fig:bl2_ma}
\end{figure}

The top panel of Figure~\ref{fig:bl2_ma} shows the pixel-pixel beam expansions,
$B_{\ell{\rm (eff)}}^2$ of equation (\ref{Bl2eff_avg}).  
The dotted lines are the results of the individual beams,
which are denoted here as $B_{\ell{\rm (i)}}^2$.  
The solid line is the result of the combined beam,
which is denoted here as $B_{\ell{\rm (c)}}^2$.  
Also plotted are
the $B_{\Sigma\ell{\rm (sm)}}^2$ (eq.~[\ref{BIsm2}])
and $B_{\Sigma\ell{\rm (ms)}}^2$ (eq.~[\ref{BIms2}]).
Here
we have used the MAXIMA-1 scans and
a pixel size of $5\times 5$ square arcminutes.
The bottom panel compares all the above $B_{\ell}^2$
to $B_{\Sigma\ell{\rm (mid)}}^2
=2/[B_{\Sigma\ell{\rm (sm)}}^{-2}+B_{\Sigma\ell{\rm (ms)}}^{-2}]$
(eq.~[\ref{B2ilmid}]),
with all line styles the same as indicated in the top panel.  

We first see that
all the $B_{\ell}^2$ have close shapes,
with considerable discrepancies only at high $\ell$
where the amplitude of $B_{\ell}^2$ is small.
Second,
the bottom panel confirms that
the $B_{\ell{\rm (c)}}^2$ is well constrained by
$B_{\Sigma\ell{\rm (sm)}}^{2}$ and $B_{\Sigma\ell{\rm (ms)}}^{2}$
(see eqs.~[\ref{B2l2_constraint}] and [\ref{Bl2eff_avg4}]),
whose fractional difference
is roughly given by ${\cal W}_\ell^2$
(see definition [\ref{W_l}]),
square of the dashed line in Figure~10.
Hence according to equation (\ref{Cl2mid_constraint})
and Figure~10,
we know that the maximum fractional error in the final $C_\ell$ estimates
by taking $B_{\ell{\rm (eff)}}^2=B_{\Sigma\ell{\rm (mid)}}^2$
for the MAXIMA-1 data
will be about ${\cal W}_\ell^2/2\approx 5\%$ for $\ell<2000$.
Although this is already a small error,
we still take $B_{\ell{\rm (eff)}}^2=B_{\ell{\rm (c)}}^2$ 
in the MAXIMA-1 data analysis for higher accuracy.
The difference between $B_{\ell{\rm (c)}}^2$ and $B_{\Sigma\ell{\rm (mid)}}^2$
is manifested by the non-zero solid line in the bottom panel of Figure~11.

In addition,
we have also verified that for both the individual and the combined beams,
the approximation  
$B_{\ell{\rm (eff)}}^2 \approx \overline{B}_{p\ell{\rm (ms)}}^2$ (eq.~[\ref{Bl2eff_avg4}])
is accurate within $1\%$ error for $\ell<2000$. 
This means that
in general situations
one can simply use $\overline{B}_{p\ell{\rm (ms)}}^2$ as the $B_{\ell{\rm (eff)}}^2$
to avoid the complicated procedure of evaluating 
the $B_{\ell{\rm (eff)}}^2$ of equation (\ref{Bl2eff_avg}) with (\ref{Bl2pp}).

\begin{figure}
  \epsscale{0.5}
  \plotone{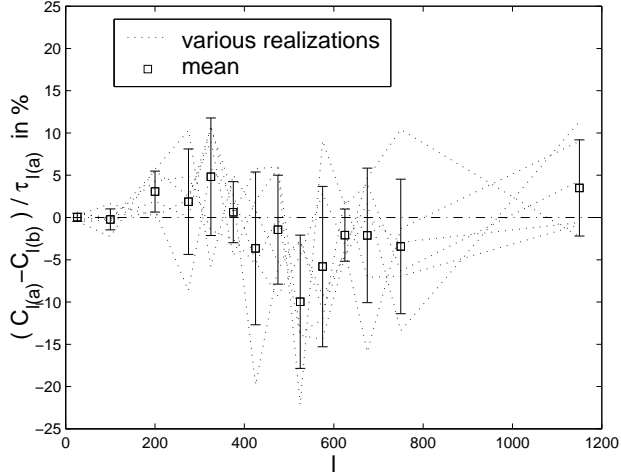}
  \caption
  {Results of simulations testing whether our formalism biases the 
  CMB angular power spectrum estimate.
  The simulations use the MAXIMA-1 scan strategy. 
  Plotted is
  the difference between angular power spectra calculated from simulations
  that have a CMB signal convolved with symmetric and  MAXIMA-1 
  (asymmetric) beams.
  The difference is normalized by the errors of one of the power spectra
  (see the text).
  The dotted lines show results
  using different realizations of the CMB signal and the noise.
  The boxes are the averages and 
  error bars are the standard deviations.}
  \label{fig:bias_test}
\end{figure}

Using equation (\ref{CBBB_app3})
with the $B_{\ell{\rm (eff)}}^2$ replaced with 
$\overline{B}_{p\ell{\rm (ms)}}^2 \Pi_{\ell{\rm (ms)}}^2$
(see eq.~[\ref{Bl2_effxapp}]), 
we tested to what extent our formalism biases the CMB angular power spectrum 
estimate. We simulated a CMB signal $\gamma_t$ in the time 
domain. Each time-domain point is allocated 
the pointing coordinates of the MAXIMA-1 scan and the signal 
is convolved with the measured MAXIMA-1 beams. In the MAXIMA-1 scan 
most pixels are scanned in two different directions. 
We then added time domain noise $n_t$ which has the MAXIMA-1 characteristics: 
an overall white noise, 
with a $1/f$ behavior at low frequencies due to the receiver response
and
a power law at high frequencies due to the electronic filtering. 
We call this ($d_t = \gamma_t + n_t$) simulation (a).
We repeated the procedure to generate simulation (b),
in which
the CMB signal is convolved with a symmetric beam 
whose power spectrum is identical to $\overline{B}_{p\ell{\rm (ms)}}^2$. 
Both simulations were then analyzed in exactly the same way,
using the procedure described in section~\ref{the_convention},
with the approximation (\ref{Bl2eff_avg4}),
(\ref{CBBB_app3}), (\ref{Bl2_effxapp}), and (\ref{Bl2_pxlapp}).
Here we have employed the quadratic estimator (Bond et al.~1998)
to estimate the power spectra $C_{\ell{\rm (a)}}$ and $C_{\ell{\rm (b)}}$
for simulations (a) and (b), respectively.
(The quadratic estimator was implemented by two
independent codes, one of which is that by Borrill (1999)
and the other by the first author,
and yielded consistent results with less than $0.1\%$ discrepancy.)
We then use 
\begin{equation}
  \label{bias}
  \lambda_\ell
  =
  \frac{C_{\ell{\rm (a)}}-C_{\ell{\rm (b)}}}{\tau_{\ell{\rm (a)}}},
\end{equation}
where $\tau_{\ell{\rm (a)}}$ is the error bar associated with
$C_{\ell{\rm (a)}}$, to quantify how much our formalism biases the
$C_\ell$ estimates.
The entire procedure is repeated six times
to yield six independent $\lambda_\ell$.
In Figure~\ref{fig:bias_test} we plot $\lambda_\ell$ Vs.\ $\ell$ for the six realizations, 
the means of these six sets of $\lambda_\ell$, and the standard deviations.
As we can see, the means 
are within $10\%$ of the error bar sizes $\tau_{\ell{\rm (a)}}$
of each $C_{\ell{\rm (a)}}$. 

With the same scan strategy, pixelization scheme, and noise property,
we repeated the same test using an extremely elliptic Gaussian beam
of 5 by 20 arcminutes in FWHM (the one we used previously).
We found again that 
the means of $\lambda_\ell$ are 
within $10\%$ of the error bar sizes $\tau_{\ell{\rm (a)}}$
for $\ell<2000$.
We therefore conclude that 
our formalism does not bias the $C_\ell$ estimates.

\begin{figure}
  \epsscale{0.5}
  \plotone{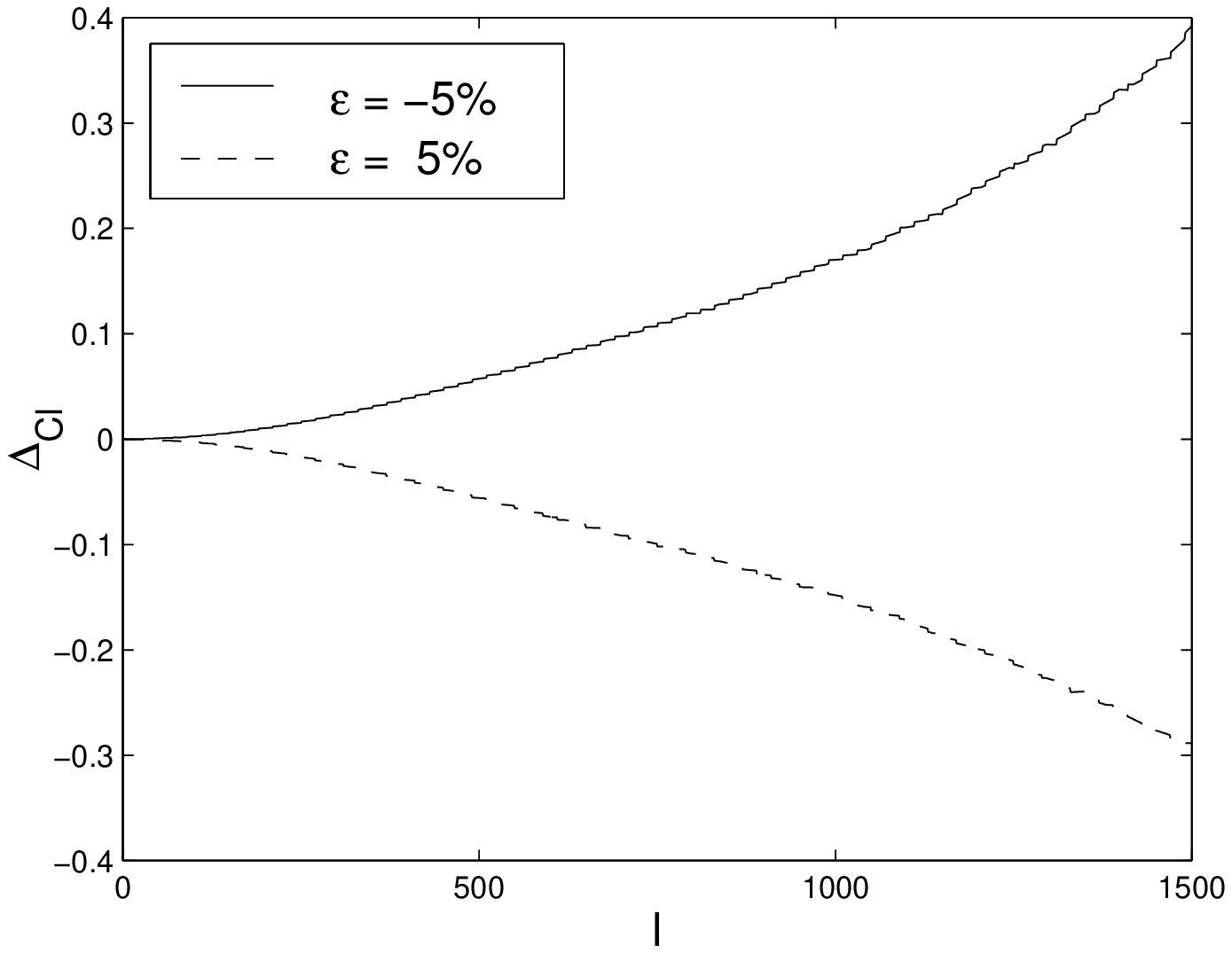}
  \caption
  {Estimated uncertainties, $\Delta_{c_\ell}=dC_\ell/C_\ell$, 
  in the $C_\ell$ estimates
  resulting from the beam measurement.}
  \label{fig:bl2err_exp}
\end{figure}

Finally, we consider the uncertainties in the $C_\ell$ estimates
resulting from uncertainties in the measurement of the beam shape,
as discussed in section~\ref{uncertainties_from_beam_measurement}.
H00 quoted an uncertainty of $\epsilon=\pm 5\%$ in the measurement 
of the MAXIMA-1 beams. The dominant contributors to this uncertainty
are of the type discussed in section \ref{uncertainties_from_beam_measurement}
(eq.~[\ref{epsilon}]) and contribute to an 
uncertainty in the $C_\ell$ estimates that is correlated between 
different $\ell$ bins.
Substituting this value into equation (\ref{Delta_Cl}) 
and using the $\overline{B}_{p\ell{\rm (ms)}}^2$ we calculated previously,
we obtain the estimated uncertainties 
$\Delta_{c_\ell}=dC_\ell/C_\ell$
in the $C_\ell$ estimates.
Figure~\ref{fig:bl2err_exp} shows the results.
As one can see,
the estimated uncertainties in the $C_\ell$ estimates are
$|\Delta_{c_\ell}| < 6\%$, $17\%$, and $40\%$
for $\ell<500$, $1000$, and $1500$ respectively.
When we include the MAXIMA-1 window-functions of $\ell$,
we find that 
for the bands used in H00, the beam size uncertainty causes less 
than $4\%$ and $11\%$ uncertainty in the 
$C_\ell$ estimates for $\ell < 410$ and $785$, respectively.

\section{DISCUSSION AND CONCLUSION}
\label{conclusion}

First we summarize the treatment of asymmetric beams 
for yielding accurate $C_\ell$ estimates
in CMB anisotropy experiments:
\begin{enumerate}
\item
Based on the measured individual beam pattern $B_0$,
one calculates the IOA, $\varpi_{0\ell}$, using equation (\ref{varpi_l}),
to quantify the level of asymmetry of the beam 
on different angular scales.
If $\varpi_{0\ell}^2/2$ is below the tolerated maximum error 
for the $C_\ell$ estimates
at the $\ell$ range of interest (see eq.~[\ref{Clmid_constraint}]),
then 
one takes $B_{\ell{\rm (eff)}}^2=B_{0\ell{\rm (mid)}}^2$ as given by equation (\ref{B20lmid})
(see also eqs. [\ref{B2lms}] and [\ref{B2lsm}] for definitions),
and goes to step~\ref{B_PI}, otherwise step~\ref{BPI}.
The resulting errors in the $C_\ell$ estimates by taking
$B_{\ell{\rm (eff)}}^2=B_{0\ell{\rm (mid)}}^2$
is quantified by equation (\ref{Clmid_constraint}).
Similarly,
when combining data from photometers of different beam shapes,
one first calculates the IOCA ${\cal W}_\ell$ using equation (\ref{W_l}),
and then employs condition (\ref{Cl2mid_constraint}) for the same check.
If ${\cal W}_\ell^2/2$ is small,
then one takes $B_{\ell{\rm (eff)}}^2=B_{\Sigma\ell{\rm (mid)}}^2$ as given by equation (\ref{B2ilmid})
and goes to step~\ref{B_PI}, otherwise step~\ref{BPI}.
\item
\label{BPI}
One checks if all the pixels have the same shape,
and if the time-stream beam $B_t$ remains unchanged 
throughout the entire observation.
If either or both of these hold, then one goes to step~\ref{B}.
Otherwise,
one calculates $B_{\Pi\ell{\rm (eff)}}^2\approx\overline{B}_{\Pi p\ell{\rm (ms)}}^2$ 
using equations (\ref{chi_t}), (\ref{mu_t}), (\ref{barBp2pi}),
 and (\ref{Bl2_effx}),
and then go to step~\ref{MAP}.
\item
\label{B}
One calculates the average pixel-beam expansion $\overline{B}_{p\ell m}$ 
using equations (\ref{barBplm2}),  (\ref{chi_t}), and (\ref{mu_t}).
We note that equation (\ref{barBplm2}) also works 
for combining data sets from
different photometers with different beam shapes,
as long as the noise level $\mu_t$ is well taken into account.
One then calculates the power spectrum $\overline{B}_{p\ell {\rm (ms)}}^2$
of $\overline{B}_{p\ell m}$ (see eq.~[\ref{Bl2eff_2}]).
This can be implemented using the form of equation (\ref{Bl2eff_3}) 
to save computation time,
i.e.\ one calculates the weighting function $f(\beta)$ first,
with discretized $\beta$, and then the $\overline{B}_{p\ell {\rm (ms)}}^2$
accordingly.
A useful check of this result is provided by
equation (\ref{B2l_constraint}) or (\ref{B2l_constraint2}).
One thus takes ${B}^2_{\ell{\rm (eff)}} \approx  \overline{B}^2_{p\ell{\rm (ms)}}$
according to equation (\ref{Bl2eff_avg4}),
and goes to the next step.
\item
\label{B_PI}
To incorporate the smoothing effect due to the pixelization of the map,
one employs equation (\ref{Bl2_effxapp2}) to obtain 
the pixel-pixel beam expansion
$B^2_{\Pi\ell{\rm (eff)}}\approx{B}^2_{\ell{\rm (eff)}}\Pi_{\ell{\rm (ms)}}^2$.
In general,
the associated $\Pi_{\ell{\rm (ms)}}^2$ can be obtained by multipole transforming 
the $\Pi({\bf x})$ that is defined in equation (\ref{Pix}).
If all pixels are regular squares,
one can instead use the convenient result in equation (\ref{Bl2_pxlapp}).
\item
\label{MAP}
One then employs equation (\ref{m}) to make a map,
and equations (\ref{L}) (or alternatives like the quadratic estimator),
(\ref{M}), (\ref{CN}), and (\ref{CBBB_app3})
to estimate the $\ell$-banded power spectrum $C_b$.
We note that in equation (\ref{CBBB_app3}),
one replaces the ${B}^2_{\ell{\rm (eff)}}$
with the $B_{\Pi\ell{\rm (eff)}}^2$ obtained previously.
\item
The uncertainties in the final band power $C_b$ 
resulting from the uncertainties
in the beam measurement can then be calculated using equation (\ref{Delta_Cl}).
In cases where the beam has a Gaussian form,
one can instead use equation (\ref{Delta_Cl_Gau}) with
the condition (\ref{lcond})
to estimate the uncertainties.
These uncertainties need to be incorporated in both
the final $C_\ell$ estimates and the estimates
of cosmological parameters.
\end{enumerate}

In previous sections,
we developed the above treatment for asymmetric beams
in order to obtain accurate $C_\ell$ estimates
at smaller angular scales.
This treatment employs the symmetric-beam approximation,
where the originally asymmetric beams are symmetrized.
The smoothing effects due to the pixelization of the CMB map are
taken into account.
The resulting uncertainties in the $C_\ell$ estimates due to 
the uncertainties in the beam measurement are also estimated.
In addition,
we derived the conditions under which
one needs to employ this formalism to account for
the asymmetry of beams.
We demonstrated certain key points
by using a simulated highly elliptic beam,
and the beams and data of the MAXIMA-1 experiment,
where the asymmetry is mild.
In particular,
we showed that in both cases 
the formalism does not bias the final $C_\ell$ estimates.

In spite of the power of the new formalism 
in dealing with various practical situations 
where the beams are not symmetric,
we should note that
it may break down under certain circumstances.
First,
if the sky patch to be analyzed has an extremely irregular shape,
then 
the important result
$  {B}^2_{\ell{\rm (eff)}}
  \approx
  \overline{B}^2_{p\ell{\rm (ms)}}
$ (eq.~[\ref{Bl2eff_avg4}])
may be invalid
due to the nonuniform distribution of $\varphi$ at each given $\Delta x$ 
(see eqs.~[\ref{Bl2eff_avg2}] and [\ref{Bl2eff_avg3}]).
Nevertheless,
the formalism as a whole is still valid in this case,
because
one can instead employ equation (\ref{Bl2eff_avg}),
$  {B}^2_{\ell{\rm (eff)}}=
    \left\langle
      {B}^2_{pp'\ell}
    \right\rangle
$,
although it is more computationally expensive.
Second,
if the total numbers of the pixels (${\cal N}_p$) 
and of the temporal samples (${\cal N}_t$) are not large,
then some statistical averages taken in the formalism
may not be appropriate (e.g., eqs.~[\ref{barBplm00}], 
[\ref{Bl2eff_avg}], [\ref{Bl2eff_avg4}], and [\ref{barBp2pi}]).
This will cause the violation of some main results
like equations (\ref{Bl2eff_avg4}), (\ref{Bl2_effx}), and (\ref{Bl2_effxapp2}).
However,
since the ${\cal N}_p$ and ${\cal N}_t$ are not large in this case,
one can always employ the full treatment of asymmetric beams
as described by equation (\ref{ClBB}).
The main results of our formalism are needed
only when ${\cal N}_p$ and ${\cal N}_t$ are large enough
to cause computational difficulty in implementing equation (\ref{ClBB}).
We note that even in the full treatment of asymmetric beams,
our results in dealing with the extra convolution effects
due to the pixelization of the map 
(see sec.~\ref{deconvolution_of_the_pixel_smoothing})
can still be employed.
Third,
the main results of our formalism have assumed
that the experimental noise in the temporal samples is independent
from each other (i.e., the white-noise assumption; see eq.~[\ref{white-noise}]),
so these results may not be suitable for experiments
that have strongly correlated noise.
Nevertheless,
as argued in the appendix,
most experiments should have only mild departure from the white noise,
and this departure does not affect our main results.
In general,
one can use condition (\ref{white-noise-cond}) or equation (\ref{dxp})
to choose a proper pixel size,
so that the white-noise approximation is still appropriate.
As we have also numerically verified,
our formalism
does not induce any bias in the final $C_\ell$ estimates
in the presence of the nonwhite noise in the MAXIMA-1 data.
Even if the experimental noise is extremely nonwhite,
we can still deal with asymmetric beams
by
employing the general results in our formalism.
This means the use of equation (\ref{barBplm0}),
together with equations (\ref{Bl2_effx}) and (\ref{CBBB_app3})
for the $C_\ell$ estimation.

In conclusion,
we have proposed a complete and well justified formalism
for the data analysis of CMB anisotropy experiments.
This formalism is very flexible and therefore 
well suited to a wide spectrum of circumstances,
especially when the experimental beams are not symmetric.
No matter how irregular the beams are,
the formalism always provides a 
both computationally economical
and statistically plausible 
way
to estimate the angular power spectrum of the CMB.
We expect this formalism to be useful not only for the small-field experiments,
but also for the full-sky experiments like PLANCK and MAP.

\acknowledgments 

JHPW and AHJ acknowledge support from 
NASA LTSA Grant no.\ NAG5-6552 and NSF KDI Grant no.\ 9872979. 
PGF acknowledges support from the RS. 
BR and CDW acknowledge support from NASA GSRP
Grants no.\ S00-GSRP-032 and S00-GSRP-031.
MAXIMA is supported by NASA Grants
NAG5-3941, NAG5-4454, by the NSF through the Center for Particle
Astrophysics at UC Berkeley, NSF cooperative agreement AST-9120005.
The data analysis used resources of the National Energy Research
Scientific Computing center which is supported by the Office of
Science of the U.S.\  Department of Energy under contract no.\
DE-AC03-76SF00098.


\appendix
\section{Non-white noise}
\label{app:colored_noise}

In this appendix, we consider the pixel-beam expansion $B_{p\ell m}$
(eq.~[\ref{BplmY0}]) 
in the case where the noise is not white (violation of eq.~[\ref{white-noise}]),
i.e., when it is correlated between pixels.
We shall show that 
even in this case,
for the purposes of determining $B_{p\ell m}$ and thus the
pixel-pixel beam expansion,
the white-noise approximation (\ref{white-noise}) is still appropriate
under certain conditions,
which are generally satisfied by practical situations.

We start with the general requirement (\ref{BplmY0}).
In the real space,
this equation is equivalent to 
\begin{equation}
  \label{BplmY0x}
  A^T_{p't'} N^{-1}_{t't} A_{tp} B_{p}({\bf x}) 
  = A^T_{p't'} N^{-1}_{t't} B_{t}({\bf x}).
\end{equation}
We note that
the center of the $B_{t}({\bf x})$ here is not at ${\bf x}=0$,
but at the location ${\bf x}_t$ of the temporal sample at time $t$.
The similar also applies to the $B_p({\bf x})$.
Thus we see that
with a given pixelization scheme (and thus given $A_{tp}$),
the relation between $B_{p}$ and $B_{t}$ depends only on
the property of $N_{t't}$.

In most experiments,
the temporal Fourier transform $\widetilde{N}(f)$ of $N_{t't}$ 
in the frequency $f$ domain
usually has the following structure:
a `$1/f$' behavior below $f_{\rm l}$ due to the receiver response,
a power law above $f_{\rm h}$ due to the electronic filtering,
and a white noise of amplitude $\mu^2$ (c.f.~eq.~[\ref{white-noise}])
 between $f_{\rm l}$ and  $f_{\rm h}$.
Since the temporal Fourier transform $\widetilde{N^{-1}}(f)$ of $N_{t't}^{-1}$
is simply the inverse of $\widetilde{N}(f)$,
we can approximate a usual $\widetilde{N^{-1}}(f)$ as
\begin{equation}
  \label{N-1f}
  \widetilde{N^{-1}}(f)
  \approx
  \mu^{-2} \left[
        H(f;f_{\rm h})-H(f;f_{\rm l})
      \right],
\end{equation}
where
$\mu^2$ is the amplitude of the white noise part 
in $\widetilde{N}(f)$,
and $H(f;f_n)$ ($n=\textrm{h, l}$) is a top-hat window function:
\begin{equation}
  \label{Hf}
  H(f;f_n)=
  \cases{
    1 & for $|f|\leq f_n$, \cr
    0 & for $|f|> f_n$,\cr
  }
  n=\textrm{h, l.}
\end{equation}
Thus in the real space we have
\begin{equation}
  \label{N-1}
  {N^{-1}_{t't}}
  \approx
  {N^{-1}_{{\rm h}t't}} - {N^{-1}_{{\rm l}t't}},
\end{equation}
where
\begin{equation}
  \label{N-1n}
  {N^{-1}_{nt't}} = \mu^{-2} \textrm{ sinc}(4\pi|t-t'|f_n), \; n=\textrm{h, l}.
\end{equation}
Here we have used the usual definition
$\textrm{sinc}(x)=2\sin(x/2)/x$.
Therefore,
to test if the white-noise approximation 
(see eqs.~[\ref{white-noise}] and [\ref{Bplmpi}]) is appropriate,
we can substitute $N^{-1}_{{\rm h}t't}$ and $N^{-1}_{{\rm l}t't}$
separately as the $N^{-1}_{t't}$ into equation (\ref{BplmY0x}),
and see if the resulting equation is consistent with the result (\ref{Bplmpi}).
We may thus derive the conditions 
under which the result (\ref{Bplmpi}) is a good approximation
even if the noise is not white.

We first consider the first term $N^{-1}_{{\rm h}t't}$ in equation (\ref{N-1}),
i.e.\ the effect from the high-frequency cut at $f_{\rm h}$.
In the white-noise case,
the first term becomes $\lim_{f_{\rm h}\rightarrow\infty}{N^{-1}_{{\rm h}t't}}$,
which is a Dirac Delta along the $t$ and $t'$ directions with a peak centered at $t=t'$.
The fact that the width of this peak is zero
means that
the correlation time of the temporal scan is zero,
so that
each sample is independent
and a pixel is related to only the temporal hits inside the pixel,
giving the form (\ref{Bplmpi}).
On the other hand,
in cases where $f_{\rm h}$ is finite,
the width of this central peak 
(the distance between the first zeros of ${N^{-1}_{{\rm h}t't}}$ 
from $t=t'$ along the $t$ or $t'$ direction)
is broadened 
(when compared to the white-noise case)
from zero to $\delta t_{\rm h}=1/f_{\rm h}$ (see eq.~[\ref{N-1n}]).
This means that
when the noise is white but with a cut-off beyond $f_{\rm h}$,
the correlation time of the temporal samples will be increased
from zero to the order of $\delta t_{\rm h}=1/f_{\rm h}$.
Therefore,
as long as the correlation time $\delta t_{\rm h}=1/f_{\rm h}$ is well below
the time required to scan across a pixel,
the pixel will have no significant correlation with the temporal hits
that are outside the pixel.
In other words,
the white-noise approximation (\ref{Bplmpi}) holds
as long as
\begin{equation}
  \label{Nhcond}
  \frac{\delta x_p}{\delta x_t} \gg \frac{\delta t_{\rm h}}{\delta t}
  = \frac{1}{f_{\rm h}\delta t},
\end{equation}
where $\delta x_p$ is the pixel size,
$\delta x_t$ is the spacing on the sky of the temporal samples,
and $\delta t$ is the integration time of each sample.
Although $\delta x_t$ is not a constant in general,
its order in a single experiment normally remains the same.

Following a similar line of logic,
we now consider the second term $N^{-1}_{{\rm l}t't}$ in equation (\ref{N-1}),
i.e.\ the effect from the low-frequency cut at $f_{\rm l}$.
In the white-noise case,
the second term becomes 
$\lim_{f_{\rm l}\rightarrow 0}{N^{-1}_{{\rm l}t't}}=\mu^{-2}$,
which is a constant along both the $t$ and $t'$ directions.
This allows us to simplify equation (\ref{BplmY0x}) as
\begin{equation}
  \label{BplmY0xw}
  \sum_p {\cal N}_{t\in p}B_p({\bf x})
  =
  \sum_t B_t({\bf x}),
\end{equation}
where ${\cal N}_{t\in p}$ is the number of temporal samples in the pixel $p$.
Thus we see that
equation (\ref{Bplmpi}) automatically fulfills the above requirement.
On the other hand,
in cases where $f_{\rm l}$ is finite,
the $N^{-1}_{{\rm l}t't}$ will remain constant along the $t$ or $t'$ direction
from $t=t'$ out to about $|t-t'|=\delta t_{\rm l}=1/2f_{\rm l}$ (the first zeros),
beyond which it begins to decay away as power law with oscillations (see eq.~[\ref{N-1n}]).
This means that
when $f_{\rm l}$ is not zero but finite,
the global requirement (\ref{BplmY0xw}) will be localized as
\begin{equation}
  \label{BplmY0xc}
  \sum_{t'\in p'} \left[\sum_{|t-t'|<\delta t_{\rm l}} B_{p\ni t}({\bf x})\right]
  \approx
  \sum_{t'\in p'} \left[\sum_{|t-t'|<\delta t_{\rm l}} B_t({\bf x})\right],
\end{equation}
where $B_{p\ni t}$ is the pixel beam of a pixel that covers ${\bf x}_t$.
In the summations over $t$ above (the summations inside the brackets),
we have ignored the contribution from $|t-t'|>\delta t_{\rm l}$
because 
 the amplitude of $N^{-1}_{{\rm l}t't}$ decays as power law
and the central (and maximum) amplitudes of the beams are always unity
(i.e.\ the contribution from $|t-t'|>\delta t_{\rm l}$ decays as a power law; 
see eq.~[\ref{N-1n}]).
Therefore,
for equation (\ref{BplmY0xc}) to hold for the white-noise result (\ref{Bplmpi}),
we require the $\delta t_{\rm l}$ to be much larger than 
the time required to scan through a pixel, i.e.\
\begin{equation}
  \label{Nlcond}
  \frac{\delta t_{\rm l}}{\delta t} 
  = \frac{1}{2f_{\rm l}\delta t}
  \gg \frac{\delta x_p}{\delta x_t }.
\end{equation}

To sum up,
we know that 
even if the noise is not white,
the white-noise approximation (\ref{Bplmpi}) is still appropriate
as long as
\begin{equation}
  \label{white-noise-cond}
  f_{\rm h}  \gg  \frac{\delta x_t}{\delta t\,\delta x_p} \gg  f_{\rm l}.
\end{equation}
In general,
the $\delta x_t$ and $\delta t$ are given by experiments,
and the $\delta x_p$ is specified by the pixelization scheme.
Therefore,
since we normally have $f_{\rm h}  \gg  f_{\rm l}$ in experiments,
condition (\ref{white-noise-cond}) can be easily satisfied
by choosing the right pixel size $\delta x_p$.
A naive choice will be
\begin{equation}
  \label{dxp}
  \delta x_p=\frac{\delta x_t}{\delta t\sqrt{f_{\rm h} f_{\rm l}}},
\end{equation}
but in fact one would usually like to choose a pixel size 
closer to the smallest limit to fully take advantage of the
experimental data.
In conclusion,
one should choose a pixelization scheme
whose pixel sizes satisfy condition (\ref{white-noise-cond}) 
(or have an order given by eq.~[\ref{dxp}]),
so that one can employ the white-noise approximation,
which leads to some main results of this paper
(see e.g., eqs.~[\ref{barBplm2}], [\ref{barBp2pi}], [\ref{Pix}],
[\ref{Bl2_effxapp2}], and [\ref{Bl2_pxlapp}]).
If condition  (\ref{white-noise-cond}) can not be fulfilled 
(i.e.\ when the noise spectrum is far from white or $f_{\rm h}  \approx  f_{\rm l}$,
which is unlikely to be the case),
one can still use the general results (\ref{barBplm0}) and (\ref{Bl2_effx})
to calculate the pixel-pixel beam expansion.





\begin{references}


\reference{BJK}
Bond, J.\ R., Jaffe, A.\ H., \& Knox L.\ 1998, \prd, 57, 2117
\reference{madcap}
Borrill, J.\ 1999, `MADCAP: The Microwave Anisotropy Dataset Computational Analysis Package'
  in ``Proceedings of the 5th European SGI/Cray MPP Workshop''
(see also astro-ph/9911389)
\reference{igloo}
Crittenden, R.\ G.\ \& Turok, N.\ G.\ 1998, \astroph{Exactly Azimuthal Pixelizations of the Sky}{9806374} 
\reference{boom} 
DeBernardis, P., et al.\ 2000, \nat, 404, 955
\reference{fj} 
Ferreira, P.\ G.\ \& Jaffe, A.\ H.\ 2000, \mnras, 312, 89
\reference{healpix}
Gorski, K.\ M., Wandelt, B.\ D., Hansen, F.\ K., Hivon, E., \& Banday, A.\ J.\ 1999, \astroph{The HEALPix Primer}{9905275}
\reference{HJS} 
Hanany, S., Jaffe, A.\ H., \& Scannapieco, E.\ 1998, \mnras, 299, 653
\reference{maxima-1} 
Hanany, S., et al.\ 2000, \apjl, accepted
\reference{tocob} 
Miller, A., et al.\ 1999, \apjl, 524, 1
\reference{tocoa} 
Torbet, E. et al.\ 1999, \apjl, 521, 79


\end{references}
\end{document}